\documentclass[10pt,conference]{IEEEtran}
\IEEEoverridecommandlockouts
\usepackage{cite}
\usepackage{amsfonts}
\usepackage{pifont}
\usepackage{enumitem}
\usepackage{makecell}
\usepackage{mathtools}
\usepackage{multirow}
\usepackage{array}
\usepackage{colortbl}
\usepackage{xcolor}
\usepackage{subfigure}
\usepackage{tcolorbox}
\usepackage{wrapfig}
\usepackage{float}
\usepackage{pict2e}
\usepackage{graphicx}
\usepackage{units}
\usepackage{tikz}

\newcolumntype{L}[1]{>{\raggedright\let\newline\\\arraybackslash\hspace{0pt}}m{#1}}
\newcolumntype{C}[1]{>{\centering\let\newline\\\arraybackslash\hspace{0pt}}m{#1}}
\newcolumntype{R}[1]{>{\raggedleft\let\newline\\\arraybackslash\hspace{0pt}}m{#1}}

\newcommand{\RNum}[1]{\uppercase\expandafter{\romannumeral #1\relax}}
\newcommand{\rNum}[1]{\expandafter{\romannumeral #1\relax}}
\newcommand{\md}[1]{{#1}}

\definecolor{g0}{RGB}{41, 205, 162}
\definecolor{g1}{RGB}{46, 238, 179}
\definecolor{g2}{RGB}{136, 252, 197}
\definecolor{g3}{RGB}{204, 255, 229}

\definecolor{gL1}{RGB}{41, 205, 162}
\definecolor{gL2}{RGB}{136, 252, 197}
\definecolor{gL3}{RGB}{204, 255, 229}
\definecolor{lgray}{gray}{0.95}

\def\BibTeX{{\rm B\kern-.05em{\sc i\kern-.025em b}\kern-.08em
    T\kern-.1667em\lower.7ex\hbox{E}\kern-.125emX}}
\begin{document}

\setlength{\abovedisplayskip}{0pt}  
\setlength{\belowdisplayskip}{0pt}  

\title{AdaptEval: A Benchmark for Evaluating Large Language Models on Code Snippet Adaptation}

%
\author{
\IEEEauthorblockN{Tanghaoran~Zhang\IEEEauthorrefmark{1}, Xinjun~Mao\IEEEauthorrefmark{1}, Shangwen~Wang\IEEEauthorrefmark{1}, Yuxin~Zhao\IEEEauthorrefmark{1}, Yao~Lu\IEEEauthorrefmark{1}, Jin~Zhang\IEEEauthorrefmark{2}, Zhang~Zhang\IEEEauthorrefmark{1}, \\ Kang~Yang\IEEEauthorrefmark{1} and Yue~Yu\IEEEauthorrefmark{1}\IEEEauthorrefmark{3}\IEEEauthorrefmark{4}\thanks{\IEEEauthorrefmark{4}Yue Yu is the corresponding author.}}
\thanks{Tanghaoran~Zhang, Xinjun~Mao, Yuxin~Zhao, Yao~Lu, Zhang~Zhang and Kang~Yang are with the State Key Laboratory of Complex \& Critical Software Environment.}
\IEEEauthorblockA{
\IEEEauthorrefmark{1}College of Computer Science and Technology, National University of Defense and Technology, Changsha, China, \\
Email: \{zhangthr, xjmao, wangshangwen13, yuxinzhao, luyao08, zhangzhang14, yangkang\}@nudt.edu.cn}
\IEEEauthorrefmark{2}Changsha University of Science and Technology, Changsha, China, Email: mail\_zhangjin@163.com \\
\IEEEauthorrefmark{3}Peng Cheng Laboratory, Shenzhen, China, Email: yuy@pcl.ac.cn
}

\maketitle

\begin{abstract}
Recent advancements in large language models (LLMs) have automated various software engineering tasks, with benchmarks emerging to evaluate their capabilities. However, for adaptation, a critical activity during code reuse, there is no benchmark to assess LLMs' performance, leaving their practical utility in this area unclear.
To fill this gap, we propose AdaptEval, a benchmark designed to evaluate LLMs on code snippet adaptation. Unlike existing benchmarks, AdaptEval incorporates \md{the following} three distinctive features:
First, \textbf{\textit{practical context}}. Tasks in AdaptEval are derived from developers' practices, preserving rich contextual information from Stack Overflow and GitHub communities.
Second, \textbf{\textit{multi-granularity annotation}}. Each task is annotated with requirements at both task and adaptation levels, supporting the evaluation of LLMs across diverse adaptation scenarios.
Third, \textbf{\textit{fine-grained evaluation}}. AdaptEval includes a two-tier testing framework combining adaptation-level and function-level tests, which enables evaluating LLMs' performance across various individual adaptations.
Based on AdaptEval, we conduct the first empirical study to evaluate six instruction-tuned LLMs and especially three reasoning LLMs on code snippet adaptation. Experimental results demonstrate that AdaptEval enables the assessment of LLMs' adaptation capabilities from various perspectives. It also provides critical insights into their current limitations, particularly their struggle to follow explicit instructions.
We hope AdaptEval can facilitate further investigation and enhancement of LLMs' capabilities in code snippet adaptation, supporting their real-world applications.
\end{abstract}


\section{Introduction}
\label{sec1:intro}
Recent advancements in large language models (LLMs) have revolutionized software development by significantly reducing the human effort in coding, ushering in the era of automatic programming~\cite{Lyu2024}. These models with billions of parameters demonstrate remarkable performance in various software engineering tasks, including code generation~\cite{Dong2024selfcollaboration,Jiang2024}, automated program repair~\cite{Xia2023,Yang2024}, \textit{etc.} To better understand LLMs' capabilities on the corresponding tasks, existing research proposes a series of benchmarks to evaluate their performance, such as HumanEval~\cite{Chen2021} and SWE-Bench~\cite{Jimenez2024}. 

Reusing code snippets is a widely adopted practice to improve the quality and efficiency of software development~\cite{Brandt2009,Gharehyazie2017,Yang2017,Wu2019,Huang2022}. A critical step in this process is adaptation, where developers modify code snippets to fit their specific context~\cite{Zhang2019,Zhang2024}. Despite its practical significance, code snippet adaptation by LLMs remains unexplored in recent research. One of the main obstacles in this domain is the lack of evaluation benchmarks.
Constructing benchmarks for adaptation, however, faces significant challenges due to three primary reasons:
\textbf{\textit{(\rNum{1}) Lack of practical context.}} Adaptation is inherently a context-aware task that involves transplanting a code snippet from one specific environment to another. Effective evaluation should therefore be grounded in practical scenarios, \textit{e.g.}, adaptation from Stack Overflow (SO) posts to developers' code bases, to accurately characterize the capabilities of LLMs in adaptation practices.
\textbf{\textit{(\rNum{2}) Lack of task requirements.}} 
Although adaptation is highly context-dependent, it is fundamentally a requirement-driven task, as its ultimate goal is to fulfill customized user needs (\textit{e.g.}, optimizing performance or extending functionality).
However, existing resources such as version control systems, often lack documentation of developers' specific adaptation requirements~\cite{Baltes2019,Zhang2024}. Hence, it remains unclear why developers make certain adaptations or what adaptation steps are necessary to integrate a code snippet into a new context. This absence of task-oriented information hinders the formulation of precise inputs for evaluating LLMs on adaptation tasks.
\textbf{\textit{(\rNum{3}) Lack of fine-grained evaluation.}} Current benchmarks primarily focus on end-to-end correctness of the LLM-generated code, only providing a binary ``correct/incorrect'' assessment. Nevertheless, an adaptation task may include a set of code edits. Evaluation on individual adaptations allows feedback on the intermediate steps, as well as LLMs' strengths and bottlenecks across diverse adaptations.

To address these challenges, we propose \textbf{\textit{AdaptEval}}, the first benchmark for code snippet adaptation, comprising \md{164 tasks with 523 adaptations in Python language}.
\md{It is designed with three distinctive features.}
\textbf{\textit{Firstly}}, each task in AdaptEval is collected from the actual adaptation practice of developers. We preserve the original context by including the referenced SO post and the associated GitHub repository for better understanding and traceability.
\textbf{\textit{Secondly}}, we annotate each task with multi-granularity descriptions: \md{Task-level annotations describe concise developer intentions, evaluating LLMs' intention understanding and code reasoning ability as intelligent assistants, while} adaptation-level ones simulate developers’ step-by-step adaptation process, serving as \md{specific instructions for LLMs to implement code changes.}
\textbf{\textit{Thirdly}}, we construct a two-tier testing framework to assess the correctness of code at both function and adaptation levels. Combined with our annotations, AdaptEval can further evaluate LLMs on individual adaptations across various types and their intermediate adaptation steps.
Overall, it takes approximately \textbf{\textit{550 person-hours}} to construct AdaptEval, covering adaptations in 38 types. Our test suite also achieves high test sufficiency with 92.95\% branch coverage and 94.38\% line coverage.

Based on AdaptEval, we first evaluate six instruction-tuned LLMs for their code snippet adaptation performance. Our results show that LLMs can solve 34.15\% to 59.15\% tasks in AdaptEval. Compared with task-level requirements, all LLMs perform significantly better with adaptation-level steps, where an increase up to 34.84\% in pass@1 is observed. Regarding adaptation types, LLMs perform best on \textit{Method Signature} and worst on \textit{Logic Customization}, with a gap of 20.31\% on average. Our analysis on their failures reveals that LLMs may violate the provided requirements due to their pre-training knowledge. Additionally, we benchmark three reasoning LLMs on AdaptEval. Results indicate that they are more effective in capturing implicit contextual cues even when no explicit requirements are provided. However, their reasoning process still need more alignments to developers’ actual adaptation strategies.

This paper makes the following contributions:
\begin{itemize}[leftmargin=*]
    \item We propose AdaptEval\footnote{ https://github.com/ztwater/AdaptEval.}, the first benchmark for code snippet adaptation, comprising 164 tasks derived from real-world, cross-platform adaptation practices of developers.
    \item \md{We introduce a distinctive evaluation design,  including multi-granularity annotation and fine-grained evaluation in AdaptEval. It supports in-depth analysis of LLMs' adaptations beyond their function-level correctness.}
    \item We conduct the first study to evaluate both instruction-tuned and reasoning LLMs on code snippet adaptation. Results suggest LLMs' strengths and limitations on adaptation tasks and point out future directions for improvements.
\end{itemize}

\begin{figure*}
    \centering
    \includegraphics[width=0.95\linewidth]{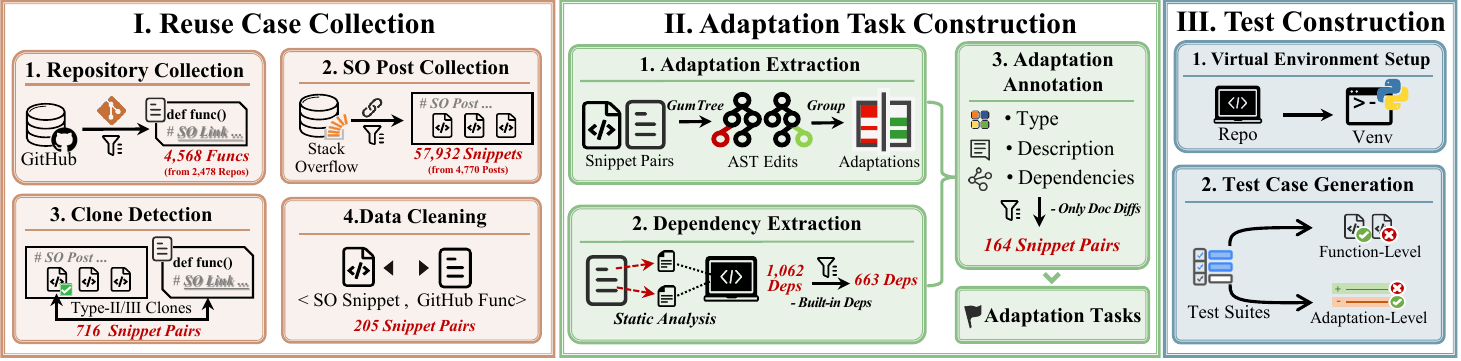}
    \caption{The construction process of AdaptEval.}
    \label{fig:framework}
\end{figure*}

\begin{figure*}
    \centering
    \includegraphics[width=0.95\linewidth]{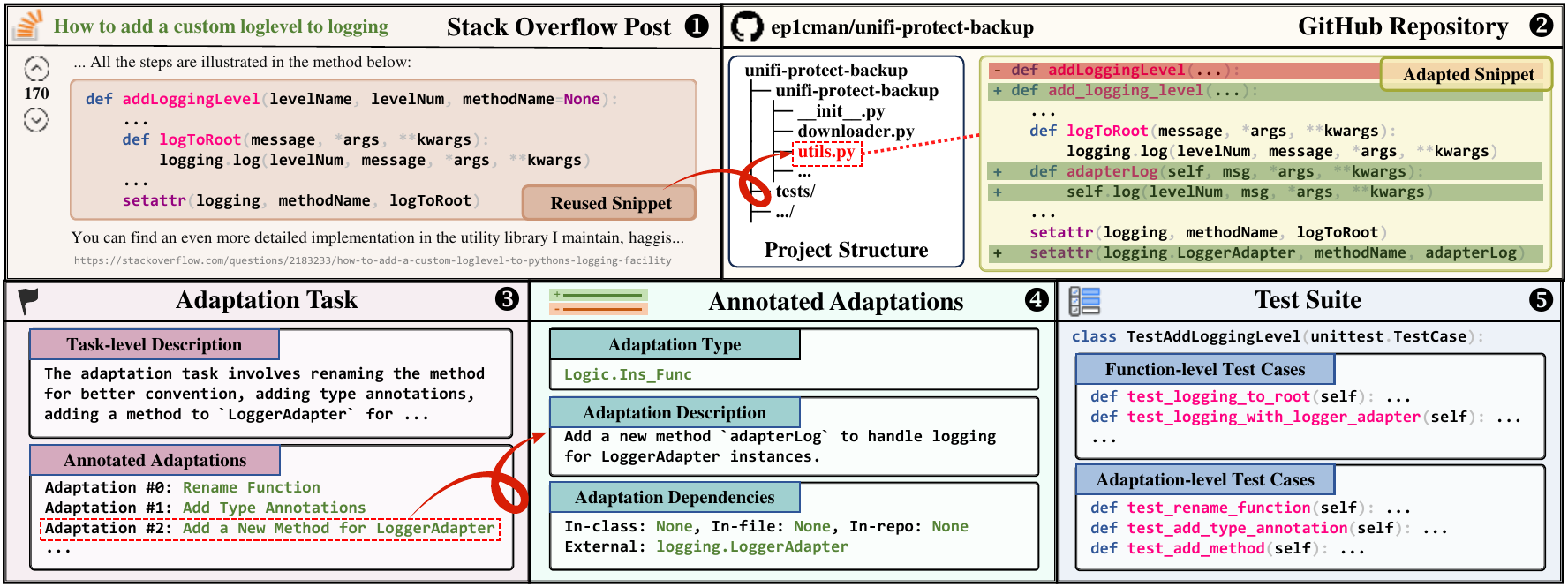}
    \caption{The overall structure of AdaptEval.}
    \label{fig:overview}
\end{figure*}

\section{Background And Related Work}
\label{sec:background}
\subsection{Code Snippet Reuse and Adaptation}
Open-source software platforms like SO provide millions of code snippets for programming problems. Reusing these solutions from the crowd benefits development efficiency and software quality. However, developers still need to adapt them to their context. Prior studies mainly focus on automating particular categories of adaptations based on historical data to reduce human efforts. Cottrell et al.~\cite{Cottrell2008} present Jigsaw to integrate snippets into developers' code, which resolve simple adaptations such as renaming. 
Zhang et al.~\cite{Zhang2019} develop a Chrome plugin, ExampleStack, to generate use templates from developers' historical adaptations.
Terragni et al.~\cite{Terragni2021} focus on a special type of adaptation, APIzation, \textit{i.e.}, transform code snippets to well-formed methods for convenient reuse. They summarize four APIzation patterns and present APIZATOR to make recommendations. 
\md{However, above adaptation tools are not effective in understanding developers' specific context and making adaptations based on it~\cite{Wu2019,Zhang2024}. Hence,} it is still challenging for developers to perform adaptations.
With the emergence of LLMs, \md{their prompt-based learning~\cite{Liu2021} ability allows them to perform unseen tasks during inference, providing opportunities for the code snippet adaptation task.
Compared to code generation, adaptation requires different abilities, including reasoning about the reuse code and modifying it to meet new constraints~\cite{Zhang2025IorI}. To advance the research and practices of LLM-based adaptation, there is a critical need for a dedicated evaluation benchmark in this domain.}

\subsection{LLMs for Code}
The emergence of LLMs brings the automation of code-related tasks to full bloom. Due to pre-training on an extremely large scale of textual and code corpora, LLMs are empowered with emerging abilities~\cite{Wei2022} on both natural language and code-related tasks. Specifically, they could be divided into three categories: (\rNum{1}) \textbf{\textit{General LLMs}} that are trained to solve a wide range of natural language tasks, such as GPT-series~\cite{OpenAI2023,OpenAI2024}, Claude-series~\cite{Anthropic2025}, and Llama-series~\cite{Meta2024}; (\rNum{2}) \textbf{\textit{Code LLMs}} that are trained specifically for code-related tasks on code corpora, such as DeepSeek-Coder~\cite{Deepseek-ai_2024b}, Codestral~\cite{MistralAI2024}, Qwen2.5-Coder~\cite{hui2024qwen2.5}; and (\rNum{3}) \textbf{\textit{Reasoning LLMs}} that are specifically designed for logical reasoning, problem-solving, and complex inference tasks, \textit{e.g.}, OpenAI's o1-series~\cite{openai2024o1}, DeepSeek-R1~\cite{deepseek-ai2025r1} and QwQ~\cite{qwq32b}. \md{However, their instruction-following abilities are observed to be degraded in recent studies~\cite{Li2025,Fu2025}. Moreover, they suffer from high inference latency and token cost, which are less practical in daily software development scenarios such as issue resolution~\cite{Jimenez2024}.}
Despite their differences, LLMs' effectiveness in code snippet adaptation has never been explored so far. 

\subsection{Existing Benchmarks for Code}

\begin{table}[ht]
    \tiny
    \centering
    \caption{The Overview of Recent LLM Benchmarks for Code.}
    \begin{tabular}{lccccccccc}
        \hline
        \textbf{Benchmark} & \textbf{Time} & \textbf{Task} & \textbf{Scale} & \textbf{Gran.} & \textbf{Dep.} & \textbf{Source} & \textbf{Test} \\
        \hline
        
        CoderEval~\cite{Yu2024} & 2023 & Gen & 230 & Func & Repo & GitHub & Func \\
        RepoBench~\cite{Liu2023} & 2023 & Gen &  26,970 & Line & Repo & GitHub & - \\
        ClassEval~\cite{Du2024} & 2023 & Gen & 100 & Class & Class & - & Class \\
        CrossCodeEval~\cite{Ding2023} & 2023 & Gen & 9,928 & Token & Repo & GitHub & - \\
        HumanEval-XL~\cite{Peng2024} & 2024 & Gen & 22,080 & Func & - & - & Func \\
        EffiBench~\cite{Huang2024} & 2024 & Gen & 1,000 & Func & - & LeetCode & Func \\
        Mercury~\cite{Du2024_Mercury} & 2024 & Gen & 256 & Func & - & LeetCode & Func \\
        EvoCodeBench~\cite{Li2024EvoCodeBench} & 2024 & Gen & 275 & Func & Repo & GitHub & Func \\
        Exec-CSN~\cite{Xie2024} & 2024 & Gen & 1,931 & Func & File & GitHub & Func \\
        DevEval~\cite{Li2024DevBench} & 2024 & Gen & 1,874 & Func & Repo & PyPI & Func \\
        OOP~\cite{Wang2024oop} & 2024 & Gen & 431& Func & Class & Multi-Src & Func \\
        ${\rm R}^2{\rm C}^2$Bench~\cite{Deng2024} & 2024 & Gen & 22,828 & Func & Repo & GitHub & - \\
        RepoClassBench~\cite{Deshpande2024} & 2024 & Gen & 287 & Class & Repo & GitHub & Class \\
        JavaBench~\cite{Cao2024JavaBench} & 2024 & Gen & 106 & Class & Repo & GitHub & Class \\
        ComplexCodeEval~\cite{Feng2024Complex} & 2024 & Gen & 11,081 & Func & Repo & GitHub & Func \\
        \hline
        SWE-Bench~\cite{Jimenez2024} & 2023 & Edit & 2,294 & Repo & Repo & GitHub & PR \\
        DebugBench~\cite{Tian2024} & 2024 & Edit & 4,253 & Func & - & LeetCode & Func \\
        CodeEditorBench~\cite{Guo2024} & 2024 & Edit & 7,941 & Func & - & Multi-Src & Func \\
        \hline
        CrossCodeBench~\cite{Niu2023} & 2023 & Multi & 54M & Func & - & Multi-Src & - \\
        XCodeEval~\cite{Khan2023} & 2023 & Multi & 25M & Prog & File & CodeForces & Prog \\
        CodeApex~\cite{Fu2024} & 2023 & Multi & 2,056 & Func & - & Internal & Func \\
        CodeScope~\cite{Yan2024} & 2023 & Multi & 13,390 & Prog & File & Multi-Src & Prog \\
        CoderUJB~\cite{Zeng2024} & 2024 & Multi & 2,239 & Func & Repo & Defects4J & Func\\
        DevBench~\cite{Li2024} & 2024 & Multi & 22 & Repo & Repo & GitHub & Repo \\
        LiveCodeBench~\cite{jain2024LiveCodeBench} & 2024 & Multi & 511 & Func & - & Multi-Src & Func \\
        \hline
        \textbf{AdaptEval} & 2025 & \textbf{Adapt} & 164 & Func & \textbf{Repo} & \textbf{SO-GitHub} & \textbf{Adapt} \\ 
        \hline
    \end{tabular}
    \label{tab:bench-list}
\end{table}

Programming and debugging are two fundamental activities in software development, which have been extensively studied~\cite{Jiang2024,Yang2023,Zhang2023Survey,Eladawy2024}. Considering their differences, we can classify all benchmarks into three categories, \textit{i.e.}, code generation, editing and multi-task. An increasing number of LLM-based approaches are adopted for them, especially after the release of ChatGPT in December, 2022. Previous benchmarks such as HumanEval~\cite{Chen2021} are considered too simple to demonstrate LLMs' capabilities. To this end, prior studies propose a series of benchmarks to better evaluate LLMs~\md{\cite{jain2024LiveCodeBench,Reux2025}}. In this paper, we collect and analyze the most recent benchmarks that (\rNum{1}) evaluate LLMs' code generation and editing abilities, and (\rNum{2}) are associated with an academic paper released between January, 2023 to October, 2024. Table~\ref{tab:bench-list} lists the overview of the 25 retrieved benchmarks. 

Existing benchmarks mainly evaluate LLMs' code generation ability, for which they are pre-trained by next token prediction paradigm~\cite{Radford2018,Radford2019,Brown2020}. Various dimensions, including the scale, efficiency and executability, are studied. Several benchmarks aim to evaluate multiple capabilities of LLMs simultaneously. For instance, CoderUJB~\cite{Zeng2024} includes code generation, test generation, program repair, and defect detection tasks. These benchmarks provide insights into LLM selection in different scenarios. Compared with generation, code editing benchmarks receive less attention. Only three benchmarks are proposed. Specifically, SWE-Bench~\cite{Jimenez2024} evaluates how LLMs can solve real GitHub issues in repo-level context. DebugBench~\cite{Tian2024} improves existing program repair benchmarks with larger-scale and richer bug types. CodeEditorBench~\cite{Guo2024} comprehensively evaluates the debugging, translating, polishing, and requirement switching tasks. However, as a significant code editing scenario, code snippet adaptation has not been supported by current benchmarks so far.

\section{The AdaptEval Benchmark}
\label{sec:adapteval}

In this section, \md{we first introduce the overall structure of AdaptEval.} Then we introduce its construction process (as shown in Figure~\ref{fig:framework}), including reuse case collection (Section~\ref{sec:collection}), adaptation task construction (Section~\ref{sec:task-construction}) and test construction (Section~\ref{sec:test-construction}). Finally, we report \md{its statistics}.

\subsection{Overall Structure}
\label{sec:overall-structure}

Figure~\ref{fig:overview} illustrates the overall structure of AdaptEval. Each task comprises three key components \md{in the following:} 
\begin{itemize}[leftmargin=*]
    \item \textbf{\textit{Practical context}}: AdaptEval captures developers' real-world reuse workflow by identifying \ding{182} the original SO post (reuse source) and \ding{183} their target GitHub repository (destination context) through mining techniques. This design ensures the veracity of our benchmark, reflecting the challenges encountered in actual reuse practices.
    \item \textbf{\textit{Multi-granularity annotation}}: AdaptEval provides adaptation requirements at two levels: \ding{184} task-level descriptions outlining \md{developer intentions}, and \ding{185} adaptation-level ones characterizing stepwise operations.
    \md{The former aims to evaluate LLMs as intelligent assistants to understand intentions and then reason for a feasible solution, while the latter evaluates LLMs as instruction-followers when actionable adaptation instructions are provided.} 
    This design aims to evaluate LLMs across diverse adaptation scenarios.
    \item \textbf{\textit{Fine-grained Evaluation}}: AdaptEval includes \ding{186} two-tier test suites with function-level and adaptation-level tests, which enables in-depth diagnosis of the adaptation process as well as LLMs' strengths and weaknesses across various adaptation types.
\end{itemize}

\subsection{Reuse Case Collection}
\label{sec:collection}
In this step, we collect real-world developers' reuse cases from two most popular communities, SO and GitHub, according to prior studies~\cite{Zhang2019,Terragni2021,Zhang2024}.
Our collection process consists of four steps. Firstly, we collect GitHub repositories and extract functions with explicit SO links. Secondly, we collect the corresponding SO posts. Thirdly, we identify candidate code reuse using a combination of explicit SO link identification and clone detection techniques to minimize spurious reuse. Finally, we clean the data by removing duplicate cases and checking the authenticity of reuse.

\subsubsection{Repository Collection}
\label{sec:reuse-repo}
We utilize GHS~\cite{Dabic2021} and GitHub REST API~\cite{GitHubAPI} to collect our GitHub repositories. The collection is based on the following criteria:
\begin{itemize}[leftmargin=*]
    \item The main language of the repository's source code is Python. The reason is that Python is one of the most popular programming languages in both SO and GitHub~\cite{Srinath2017}, which contains adequate data for benchmark construction.
    \item The repository is non-forked and was created after 2023. 
    This rule aims to avoid repetition and \md{balance the dataset size and potential data leakage risk.}
    \item At least one Python code file in the repository has an explicit reference to a SO post within the method scope. This rule filters out reuse cases that do not target coding purposes.
\end{itemize}
We identify a total of 2,478 Python repositories. To isolate reused code from later evolutionary changes, we follow Zhang et al.~\cite{Zhang2024} to collect the file version immediately after the commit that introduced the code. This step filters out non-adaptation changes. Besides, we also preserve all associated commits as authentic process records for subsequent annotation. Finally, we extract 4,568 functions with SO links.

\subsubsection{SO Post Collection} 
To retrieve reused SO snippets, we use the Stack Exchange API~\cite{StackAPI} to collect SO posts according to the reference links in the GitHub source code files. We observe that developers may refer to either the question post or a specified answer post in their source code files. To obtain the complete reuse context, we collect the whole SO question post instead of a single answer. Specifically, we first retrieve the corresponding parent question posts for each linked answer. Then, for each question post, we query its \textit{id}, \textit{title}, \textit{body}, \textit{answers}, \textit{owner}, \textit{tags}, \textit{score}, \textit{comments} fields and timestamps for activities. For each answer belonging to the question, we collect its \textit{id}, \textit{body}, \textit{owner}, \textit{is\_accepted}, \textit{score} fields and timestamps for activities. Finally, we extract all code spans as individual code snippets from each SO answer post, serving as potential reused snippets. 
\md{Followed Zhang et al.~\cite{Zhang2024},} we discard posts with non-positive scores \md{to consider both the quality and coverage of our data}, resulting in 57,932 snippets from 4,770 posts eventually.

\subsubsection{Clone Detection}
To identify the candidates of code reuse, prior studies often employ two approaches including link identification~\cite{Terragni2021,Zhang2024} and code clone detection~\cite{Wu2019,Zhang2019,Mondal2019,Huang2022}. The former identifies explicit SO references in GitHub snippets but could include spurious reuse, such as references to problems or explanations. The latter ensures the code similarity but may introduce false positives, \textit{e.g.}, coincidental clones, trivial snippets, \textit{etc}. Following Terragni et al.~\cite{Terragni2021}, we combine the clone detection technique and explicit SO link references to obtain candidate code reuse, which are more likely to be genuine. 

We first pair each \md{collected} function from GitHub with SO snippets extracted from the linked post to avoid identifying coincidental clones from other posts. For each $\langle\text{GitHub function}, [\text{SO snippet list}]\rangle$ pairs, we use SourcererCC~\cite{Sajnani2016} to search Type-\RNum{2} and Type-\RNum{3} clones as we are only interested in explicit code reuse with adaptations. Type-\RNum{1} clones only include differences in blank spaces and comments, which does not require any adaptations, while Type-\RNum{4} clones (semantically equivalent but with little syntactical similarity) may not suggest explicit reuse and are difficult to detect. Aligned with previous studies~\cite{Zhang2019,Terragni2021}, we determine two snippets as clones by setting the token similarity threshold to 70\% for the best tradeoff between the precision and recall. We discard all functions without any clones in their linked posts as spurious reuse. As for functions with multiple clones from SO, we select the one with the highest similarity. After clone detection, we obtain 716 snippet pairs.

\subsubsection{Data Cleaning}
\md{Different developers may perform exactly the same adaptations accidentally.} Such duplicate data in our benchmark may lead to unfair evaluation. To this end, we exclude snippet pairs whose source SO snippet and GitHub function are both identical, \md{resulting in 232 snippet pairs.} 
Although we employ filtration rules and clone detection techniques to identify potential code reuse, there could still be exceptions, \textit{e.g.}, code written in another programming language in SO answer posts. We further filter these cases manually. Eventually, we obtain 205 snippet pairs.

\subsection{Adaptation Task Construction}
\label{sec:task-construction}
As developers' original adaptation requirements are not properly recorded, we aim to reconstruct adaptation tasks from their actual practices. To accurately describe them, we construct each task following a bottom-up manner. Specifically, we first utilize \textit{GumTree}~\cite{Falleri2014} to extract code edits at the Abstract Syntax Tree (AST) level and group them into 
\md{\textit{\textbf{adaptations}}, which serves as the unit of evaluation in AdaptEval.}
Subsequently, we develop a static analysis tool with \textit{tree-sitter}~\cite{TreeSitter} to extract their dependencies as relevant context. Then we annotate each adaptation with a detailed description, a type label and required dependencies. Finally, we combine all annotated adaptations in a single reuse case as an adaptation \textbf{\textit{task}}, and summarize a task-level requirement.

\subsubsection{Adaptation Extraction}
\label{sec:adaptation-extraction}
In this step, we first extract all code edits from reuse snippet pairs leveraging a popular code-differencing tool, \textit{GumTree}~\cite{Falleri2014}.
However, individual AST-level edits are too fine-grained for understanding and description~\cite{Huang2018}. For instance, inserting an \textit{if} statement to handle a new condition may include numerous AST edits, which should be considered as a whole. To better reflect developers' intentions, we group the edits according to their syntax adjacency on the AST following CLDiff~\cite{Huang2018}.

\subsubsection{Dependency Extraction}
\label{sec:dep-extraction}
Since adaptation requires developers to integrate code snippets into their code bases, we extract all repository-level context dependencies for each adapted function in AdaptEval, including four types, \textit{i.e.}, intra-class, intra-file, intra-repo and external.
For each adapted function, we first extract all its used identifiers from its AST. Then we extract defined classes, functions, and global variables from the current file containing the adapted function. If an identifier is identical to a class, field or method name, it belongs to the intra-class dependency, denoted as ``\textit{ClassA.MethodB}''. If an identifier is defined as a global variable or a function, then it belongs to the intra-file dependency, simply denoted as their name.
Subsequently, we distinguish intra-repo and external dependencies by parsing the import statements and the repository files. If an identifier is located within the repository, it belongs to the intra-repo dependencies, denoted by ``\textit{Path/To/File::ClassA.MethodB}''.
Identifiers with imported third-party library names belong to external dependencies, denoted by their full-qualified names without alias, \textit{e.g.}, ``\textit{numpy.array}''. The identifiers of builtin functions, variables and constants in Python 3 are discarded.

\subsubsection{Adaptation Annotation}
During this step, we manually annotate each adaptation with its type and description. This process is carried out by five participants with at least three years of Python programming experience. Two of them are responsible for writing the adaptation description, while another two conduct the open-coding process of adaptation types. Each member in both groups takes charge of half the cases and then double-checks the other's annotation for cross-validation. The remaining participant with the most experience serves as the mediator and final reviewer. 
We first annotate adaptation descriptions, \md{whose criteria include: intent-based, concise and consistent.} 
To ensure that they accurately reflect developers' intents, we employ a rigorous and multi-step protocol. Before annotation, we conduct a pilot study with annotators as well as two senior engineers with industrial experience on five real-world examples. All participants are provided with: (1) the original SO post, (2) the integration commit (Section~\ref{sec:reuse-repo}), and (3) their linked artifacts (commit messages, discussions) to extract the rationale behind each adaptation. All participants found the integration commit helped them comprehend the original intent, especially when commit messages and inline comments reveal the motivations. 
To this end, we require annotators carefully refer to them during annotation. \md{During the discussion, we set a 50-word threshold for each description, which could balance the conciseness and detail.}
Beside, we share an online document with well-designed demonstrations and dynamically update it to ensure consistency throughout the annotation. To mitigate subjectivity, our cross-validation requires the mediator to resolve every disagreement through discussion until consensus is achieved. Our approach minimizes individual bias while aligning our annotations with accessible records documented in commit histories.
After all descriptions are completed, the other group annotates adaptation types following the principles of thematic analysis~\cite{Braun2006}. The participants generate the initial codes using a \textit{verb-object} phrase, \textit{e.g.}, Rename Function, to characterize the adaptation. After cross-validation, they work together to group them into themes. Then they iteratively re-evaluate and group the themes until they are established. The final themes are reviewed by the final reviewer. If there are any inappropriate expressions, the three participants will discuss and refine them to reach the consensus. 
\md{The Cohen's Kappa value~\cite{Cohen1960} between two annotators during the cross-validation phase is 0.929, while the Cohen's Kappa value between annotators and the final reviewer on the final themes is 0.975.}
Finally, we obtain descriptions for each adaptation and the adaptation taxonomy.

As adaptations in docstrings or comments do not affect the functionality, we exclude them as well as the cases with only adaptations of this type from AdaptEval. It allows us to focus on significant adaptations, thereby reducing noise in the benchmark. The derived adaptation taxonomy is shown in Table~\ref{tab:taxonomy}. 
\md{Specifically, it includes three categories, \textit{Method Signature}, \textit{Logic Customization} and \textit{Refactoring}, with 128, 227, and 168 adaptations respectively. \textit{Method Signature} refers to adaptations made to the method interface while preserving the logic in its body. \textit{Logic Customization} refers to the functional changes in the method body. \textit{Refactoring} refers to non-functional adaptations which do not change the program behavior.}
The most prevalent adaptation is \textit{Rename Function}, with 70 occurrences. It indicates that developers often need a more descriptive or consistent function name during adaptation. 
\md{While Zhang et al. propose an insightful taxonomy~\cite{Zhang2019} for adaptation, it is based on Java and does not consider the recent studies on APIzation~\cite{Terragni2021}, which is revealed by our \textit{Method Signature} category.}

\begin{table}[ht]
    \scriptsize
    \centering
    \caption{The Adaptation Taxonomy on AdaptEval}
    \begin{tabular}{ll}
        \hline
        \textbf{Category} & \multicolumn{1}{c}{\textbf{Adaptation Type}}\\
        \hline
        MS (128) &
        \makecell[l]{Encapsulate (13), Rename Function (70), Update Function Type (11),\\Insert/Delete/Update Parameter (13/5/16)} \\
        \hline
        LC (227) &
        \makecell[l]{Initialize/Replace Variable (3/11), Update Constant (13), Convert Ob-\\ject Type (23), Insert/Delete/Update Call (22/8/52), Insert Import\\ (16), Insert/Delete/Update Return (5/1/8), Insert/Delete/Update Cond-\\ition(31/13/9), Handle Exception (11), Insert Function (1)}\\
        \hline
        RE (168) &\makecell[l]{Insert/Delete/Update Type Annotation (52/3/1), Insert/Delete Tempor-\\ary Variable (9/3), Rename Parameter/Variable (24/20), Update API\\Referenced Name (25), Insert None Return (2), Refactor/Move Exp-\\ression (17/1), Delete Import (3), Split Lines (2), Inline/Expand/\\Move Function (1/3/2)} \\
        \hline
    \end{tabular}
    \label{tab:taxonomy}
\end{table}

Based on our extracted dependencies, we associate each adaptation with its dependencies by determining whether its updated code elements are present in our dependency set. To obtain task-level requirement description, we first group all adaptations by their types as they are likely for the same purpose. Then we summarize each group into a concise phrase and combine them as a one-sentence description. This ensures that the task-level descriptions are both comprehensive and concise. Finally, we derive our final set of 164 adaptation tasks.

\subsection{Test Construction}
\label{sec:test-construction}

\begin{figure*}[htbp]
    \centering
    \includegraphics[width=\linewidth]{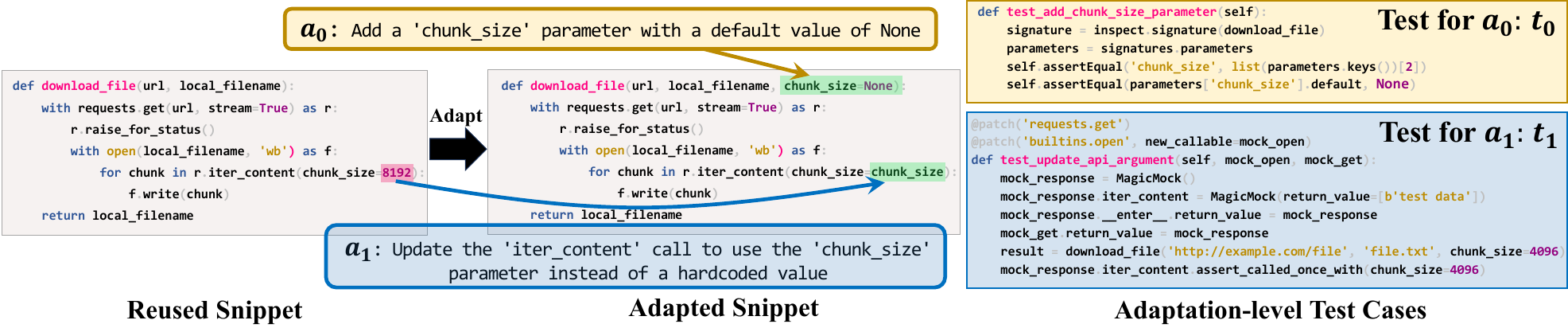}
    \caption{\md{An example of adaptation-level test cases in AdaptEval.}}
    \label{fig:test-construction}
\end{figure*}

As our benchmark needs to support an automatic evaluation of LLM-adapted code from hundreds of repositories, we manually set up a virtual testing environment and create a test suite for each task in AdaptEval. Apart from the overall correctness of the adapted function, we also enable a fine-grained evaluation using adaptation-level test cases. In the testing pipeline, we automatically extract the adapted function from LLMs' output, inject the code under test into the corresponding context, and run the tests in a virtual environment.

\subsubsection{Virtual Environment Setup}
To avoid dependency conflicts across different repositories, we deploy each repository in an individual virtual environment and install its required dependencies.
Specifically, we first look for the supported Python version specified in the documentation of each repository. Then we use \textit{venv} to create a virtual environment in its root directory and use \textit{pip} to install all the required Python packages specified in the dependency list.
To alleviate the burden of deployment, we only install the necessary dependencies that allow the execution of the adapted functions.

\subsubsection{Test Case Generation}
We construct two-tier test suites containing both function and adaptation-level test cases in AdaptEval to evaluate LLMs' adaptations.
Function-level test cases assess the overall functionality of the adapted code. They ensure that functional adaptations correctly modify program behaviors while preserving those of the remaining code.
However, only considering overall functionality is insufficient. First, it cannot evaluate individual adaptations, as LLMs may partially succeed in some adaptations yet fail function-level tests. Fine-grained tests are essential for identifying LLMs' effectiveness and bottlenecks across in different adaptation types. Second, refactoring, which is also critical in software reuse, cannot be directly evaluated through program behavior alone. Therefore, AdaptEval incorporates adaptation-level test cases for a more comprehensive evaluation.

\md{The core criterion for our adaptation-level tests is their discriminative power: each test should pass only when its target adaptation is correctly applied and fail otherwise.} Specifically, given a sequence of adaptations $\mathcal{A}=\{a_0, a_1, \ldots, a_n\}$, our objective is to write adaptation-level tests $\mathcal{T}=\{t_0, t_1, \ldots, t_n\}$ such that passing $t_i$ ensures that the corresponding adaptation $a_i$ is made to the reused snippet and is not influenced by subsequent adaptations $\{a_{i+1}, \ldots, a_n\}$. 
For instance, Figure~\ref{fig:test-construction} illustrates an example of adaptation-level test cases in AdaptEval. $a_0$ is a default parameter addition and $a_1$ updates an API call with the added argument. We construct $t_0$ to validate that the parameter is added in the correct place and its default value is \textit{None}. For $a_1$, we write $t_1$ to call the adapted function with a different \textit{chunk\_size} value, \textit{4096}, and use mock objects to determine whether the \textit{iter\_content} method is called with that value precisely.

We consider reusing the original tests written by developers for our adapted functions. However, 142 out of 164 tasks lack unit tests in their repositories and the remaining tests are insufficient. 
Hence, we manually construct function-level and adaptation-level test cases for each task, ensuring their discriminative power and coverage. The construction process involves four participants with at least three years of Python programming experience.
Three of them are responsible for writing test cases while the last deploys all repositories and rigorously checks the test cases to ensure their correctness and quality. The test suites in AdaptEval are written using the Python \textit{unittest} framework. 
The idea of ``mocking''~\cite{Vladimir2020} is also adopted to improve the stability of the test and provide a controlled and efficient testing environment. We try our best to obtain a line-level and branch-level coverage of 92.95\% and 94.38\%. Each adaptation is covered by at least one test case to validate whether it is performed correctly as required.

\subsection{Dataset Statistics}
\label{sec:stats}
\begin{table}[htbp]
    \footnotesize
    \centering
    \caption{The Overall Statistics of Tasks in AdaptEval.}
    \begin{tabular}{lrrr}
        \hline
        \textbf{Metric} & \textbf{Mean} & \textbf{Total} \\
        \hline
        Line-of-Code (LOC)  & 18.9 & 3,106 \\
        Num of Adaptations  &  3.2 &   523 \\
        Num of AST Edits    & 13.6 & 2,236 \\
        Description Length  & 14.6 & 2,389 \\
        Num of Dependencies &  4.0 &   663 \\
        Num of Tests        &  6.6 & 1,079 \\   
        \hline
    \end{tabular}
    \label{tab:dataset-scale}
\end{table}

Our AdaptEval benchmark has 164 adaptation tasks with 523 adaptations. As illustrated in Table~\ref{tab:dataset-scale}, the average line-of-code (LOC) of each task is 18.9. In each task, LLMs are required to perform 3.2 \md{adaptations} on average, while the average number of corresponding AST edits is 13.6. It demonstrates that our \md{AdaptEval} is composed of concise and higher-level adaptations. The average length of adaptation-level requirements is 14.6 words. Each task has 4.0 dependencies and is equipped with 6.6 test cases on average.

\section{Evaluation Setup}
\label{sec:eval-setup}
This section describes our evaluation setup to explore LLMs' adaptation ability based on AdaptEval's unique features. 

\subsection{Research Questions}
We aim to \md{benchmark LLMs' adaptation capabilities by addressing the following four research questions as follows:} 

\begin{itemize}[leftmargin=*]
    \item \textbf{RQ1 \md{(Task-Level Performance)}: How do LLMs perform on code snippet adaptation tasks \md{across diverse scenarios}?} \md{We investigate LLMs' performance in solving whole adaptation tasks on AdaptEval with different settings.}
    \item \textbf{RQ2 \md{(Adaptation-Level Performance)}: How do LLMs perform in making different types of adaptations?} 
    \md{To investigate current bottlenecks of LLMs, we investigate their adaptation-level performance across different types.} 
    \item \textbf{RQ3 (Error Analysis): What are the prevalent errors made \md{in LLMs' adaptations}?}
    \md{We further analyze LLMs' failed adaptations to better understand their limitations.}
    \item \textbf{RQ4 (Potentials of Reasoning LLMs): What are the potentials of reasoning LLMs in adaptation?} 
    \md{Due to the distinct capabilities of reasoning LLMs, we separately evaluate their performance in a more challenging setting.}
\end{itemize}

\subsection{Studied Models}
As the adaptation task requires both advanced coding and instruction-following abilities, we first select six state-of-the-art instruction-tuned LLMs. Particularly, we adopt four \textbf{general LLMs} for their capabilities in both natural language and coding tasks: GPT-4o (2024-11-20)~\cite{OpenAI2024}, DeepSeek-V3 (2024-12-26)~\cite{deepseek-ai2024v3}, Gemini-2.0-Flash (2025-02-05)~\cite{Google2025} and Llama-3.3-70B (2024-12-06)~\cite{Meta2024};
and two \textbf{code LLMs}: Qwen2.5-Coder-14B-Instruct (2024-09-19)~\cite{hui2024qwen2.5} and Codestral-22B (2024-05-29)~\cite{MistralAI2024}, for their code-specific ability.
To further explore LLMs' potentials in inferring developers' adaptation intents, we also include three \textbf{reasoning LLMs}, DeepSeek-R1 (2025-01-20)~\cite{deepseek-ai2025r1}, Claude-3.7-Sonnet (2025-02-19)~\cite{Anthropic2025} and QwQ-32B (2025-03-06)~\cite{qwq32b} in our experiments.
Our selected LLMs cover both closed- and open-source ones in leading series, all released since 2024, which ensures a comprehensive and timely evaluation of their adaptation ability.

\subsection{Adaptation Settings}
Since adaptation is a requirement-driven and context-dependent task, we propose a two-dimensional setting to evaluate LLMs' adaptation performance. First, we design three requirement settings, including \textbf{AReq}, \textbf{TReq} and \textbf{NoReq} to evaluate the influence of different instruction granularities. 
\textbf{AReq} provides LLMs with our specific adaptation-level instructions \md{for each code change}, while \textbf{TReq} offers only task-level requirement \md{outlining developer intentions}.
To demonstrate the capabilities of reasoning LLMs, we introduce a more challenging setting, \textbf{NoReq} in RQ4, where LLMs must infer developer intentions from intra-file context and perform adaptation end-to-end without explicit requirements.
As for the context setting, we consider two scenarios, \textit{Oracle} and \textit{NoCtx}, to assess the role of contextual dependencies. \textit{Oracle} provides LLMs with all dependent context, simulating an ideal scenario with prior knowledge of dependencies. In contrast, \textit{NoCtx} requires LLMs to adapt solely based on the given requirements, reflecting a more constrained setting.

\subsection{Prompt Design}
\begin{figure}
    \centering
    \includegraphics[width=0.9\linewidth]{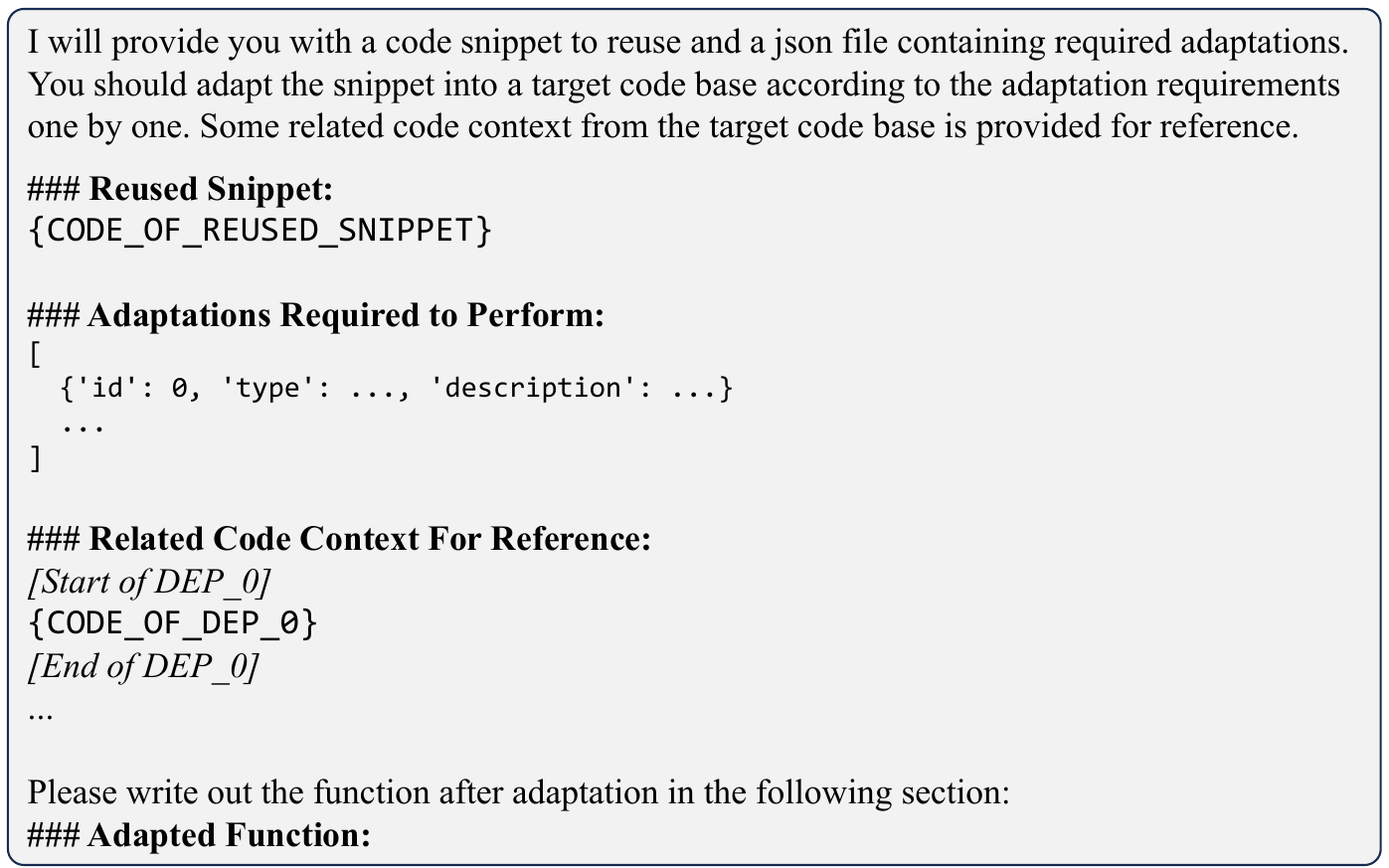}
    \caption{The prompt template for AdaptEval's evaluation (\textbf{AReq} \& \textit{Oracle}).}
    \label{fig:prompt}
\end{figure}

We design \md{a set of prompts to support LLMs' adaptations in different settings.} Figure~\ref{fig:prompt} shows the prompt template in the most informative setting (\textbf{AReq} and \textit{Oracle}). The prompt begins with a \md{brief description of the task}. The first block provides the reused snippet as input, while the second block specifies our adaptation-level requirements step by step. The third block includes all dependent contexts. Finally, it offers an instruction for LLMs to generate the adapted function as output. Prompts for \textbf{TReq} and \textbf{NoReq} replace the second block with task-level instructions or intra-file context, respectively, while the \textit{NoCtx} prompt removes the third block.

\subsection{Evaluation Metrics}
We adopt an execution-based metric pass@$k$ to check the correctness of LLMs' adaptations. It measures the likelihood that a LLM produces a correct solution within $k$ attempts based on unit test execution.
Compared to similarity-based metrics, execution-based metrics can measure the functional correctness of the outputs with higher reliability~\cite{Chen2021}.

In AdaptEval, we are interested in both the overall correctness of function-level adaptation task and individual adaptations, so we calculate both pass@$k$-t and pass@$k$-a. Specifically, if an adapted function passes all test cases of a task/adaptation, it will be considered as a correct sample for the task/adaptation. 
Considering the generation cost, we randomly generate five responses ($n=5$) in line with previous work~\cite{Du2024}. For task-level evaluation, we report pass@$k$ metrics with $k\in\{1,5\}$, reflecting LLMs' accuracy with effort control in practical use, \textit{i.e.}, whether a correct adaptation can be made through at most five trials. For adaptation-level evaluation, we report the pass@1-a to solely evaluate their correctness.

\subsection{Implementation Details}
As LLMs are non-deterministic models, different settings for their randomness may lead to quite different results~\cite{Ouyang2023,Doderlein2023}. In line with previous work~\cite{Du2024,Cao2024JavaBench}, we use nucleus sampling~\cite{Holtzman2019} to randomly generate five solutions with a temperature of 0.2 for all LLMs. All experiments are conducted with two GeForce RTX 4090-24G GPUs on Ubuntu 22.04.

\section{Results}

\subsection{RQ1: \md{Task-Level} Performance}
\begin{table}[htbp]
    \scriptsize
    \centering
    \caption{The Task-Level Performance of Six Instruction-tuned LLMs in Different Settings of AdaptEval.}
    \begin{tabular}{l|c|cc|cc}
        \hline
        \multirow{2}{*}{\textbf{Model}} & \multirow{2}{*}{\textbf{Context}}  & \multicolumn{2}{c|}{\textbf{TReq}} & \multicolumn{2}{c}{\textbf{AReq}} \\
        \cline{3-4}\cline{5-6}
        & & pass@1-t & pass@5-t &  pass@1-t & pass@5-t \\
        \hline
        \multirow{2}{*}{GPT-4o}
        & \textit{NoCtx}  & \textbf{16.46} & \textbf{17.07} & 52.20 & \textbf{57.93} \\
        & \textit{Oracle} & \textbf{20.73} & \textbf{23.17} & \textbf{59.15} & \textbf{63.20} \\
        \hline
        \multirow{2}{*}{DeepSeek-V3} 
        & \textit{NoCtx}  & 15.85 & \textbf{17.07} & \textbf{54.76} & 57.32 \\
        & \textit{Oracle} & 19.39 & 20.12 & 57.31 & 60.37 \\ 
        \hline
        \multirow{2}{*}{Gemini-2.0} 
        & \textit{NoCtx}  & 13.78 & 14.02 & 52.80 & 54.27\\
        & \textit{Oracle} & 15.85 & 17.68 & 54.63 & 57.32 \\
        \hline
        \multirow{2}{*}{Llama-3.3} 
        & \textit{NoCtx}  & 8.54 & 10.36 & 29.27 & 33.53\\
        & \textit{Oracle} & 8.54 & 12.80 & 34.15 & 43.29\\
        \hline
        \multirow{2}{*}{Qwen2.5-Coder}
        & \textit{NoCtx} & 12.44 & 13.41 & 44.02 & 48.78 \\
        & \textit{Oracle} & 16.83 & 18.90 & 48.66 & 53.05 \\
        \hline
        \multirow{2}{*}{Codestral}
        & \textit{NoCtx} & 11.58 & 12.19 & 46.71 & 51.22 \\
        & \textit{Oracle} & 14.51 & 17.07 & 50.98 & 54.88  \\
        \hline
    \end{tabular}
    \label{tab:res-overall}
\end{table}

Table~\ref{tab:res-overall} shows the overall performance of six instruction-tuned LLMs. We use bold text to highlight the best pass@1-t and pass@5-t scores in each setting. Under the most informative setting (\textbf{AReq} \& \textit{Oracle}), we can observe that instruction-tuned LLMs can solve 34.15\% to 59.15\% tasks in AdaptEval. The closed-source GPT-4o achieves the best pass@1-t and pass@5-t of 59.15\% and 63.20\%. The second-tier LLMs include DeepSeek-V3 and Gemini-2.0, whose pass@1-t lag behind GPT-4o by 3.11\% and 7.64\%. Llama-3.3 ranks last among all LLMs, whose pass@1-t is 42.27\% less than GPT-4o. We observe that code LLMs are inferior to the state-of-the-art general LLMs across all settings. This suggests that the instruction-following ability is more important in adaptation and simply pre-training LLMs' on code corpora does not necessarily improve their performance.

Compared with task-level requirements (\textbf{TReq}), our adaptation-level instructions (\textbf{AReq}) demonstrate a substantial improvement, with an average increase of 34.84\% in pass@1-t. The relative growth of pass@1-t ranges from 185\% to 299\%. 
\md{It implies that LLMs still struggle to reason for a complete adaptation solution from developer intentions only, but they could benefit from clear and actionable adaptation instructions.}
In the \textit{Oracle} setting, most LLMs exhibit a 2\% to 7\% absolute improvement compared to the \textit{NoCtx} setting. It indicates that retrieved context includes helpful information for adaptation. 

\begin{center}
    \begin{tcolorbox}[colback=lgray,colframe=black,width=\linewidth,arc=1mm,boxrule=1pt,top=2pt,bottom=2pt,left=3pt,right=3pt]
        \textbf{Finding 1:} Instruction-tuned LLMs can solve up to 59.15\% tasks in AdaptEval. Compared with task-level requirements, our annotated adaptation steps significantly improve LLMs' performance with a rise of 34.84\% in pass@1-t. 
    \end{tcolorbox}
\end{center}

\subsection{RQ2: \md{Adaptation-Level Performance}}

\begin{table}[htbp]
    \tiny
    \centering
    \caption{\md{LLMs' Adaptation-Level Performance Across Different Types.}}
    \begin{tabular}{l|L{45pt}|C{16.4pt}C{16.4pt}C{16.4pt}C{16.4pt}C{16.4pt}C{16.4pt}}
        \hline
        \multicolumn{1}{c}{\textbf{Cat.}} & \multicolumn{1}{|c|}{\textbf{Type}} & GPT-4o & DS-V3 & Gemini & Llama & Qwen & Codestral \\
        \hline
        \multirow{8}{*}{MS}
        & Encap (13)
        & \cellcolor{gL2}\ \;\textbf{8.0} & \cellcolor{gL2}\ \;\textbf{8.0} & \cellcolor{gL2}\ \;\textbf{8.0} & \cellcolor{gL3}\ \;4.0 & \cellcolor{gL2}\ \;7.4 & \cellcolor{gL2}\ \;\textbf{8.0} \\
        & Rnm\_Func (70)
        & \cellcolor{gL1}\textbf{67.6} & \cellcolor{gL1}64.0 & \cellcolor{gL1}66.8 & \cellcolor{gL3}32.0 & \cellcolor{gL1}65.4 & \cellcolor{gL1}64.2 \\
        & Upd\_Type (11)
        & \cellcolor{gL1}\ \;9.6 & \cellcolor{gL1}\textbf{10.0} & \cellcolor{gL1}\textbf{10.0} & \cellcolor{gL3}\ \;4.0 & \cellcolor{gL1}\ \;9.8 & \cellcolor{gL1}\ \;9.0 \\
        & Ins\_Param (13)
        & \cellcolor{gL2}\ \;9.0 & \cellcolor{gL2}\ \;9.0 & \cellcolor{gL2}\ \;8.0 & \cellcolor{gL2}\ \;7.0 & \cellcolor{gL2}\ \;9.0 & \cellcolor{gL2}\ \;\textbf{9.4} \\
        & Del\_Param (5)
        & \cellcolor{gL1}\ \;4.0 & \cellcolor{gL1}\ \;4.0 & \cellcolor{gL1}\ \;4.0 & \cellcolor{gL2}\ \;3.0 & \cellcolor{gL1}\ \;4.0 & \cellcolor{gL1}\ \;\textbf{4.2} \\
        & Upd\_Param (16)
        & \cellcolor{gL1}\textbf{13.0} & \cellcolor{gL1}\textbf{13.0} & \cellcolor{gL1}12.0 & \cellcolor{gL3}\ \;7.0 & \cellcolor{gL1}\textbf{13.0} & \cellcolor{gL1}12.4 \\
        \cline{2-8}
        & \textbf{All Types (128)}
        & \cellcolor{gL1}\textbf{111.2} & \cellcolor{gL1}108.0 & \cellcolor{gL1}108.8 & \cellcolor{gL3}\ \;57.0 & \cellcolor{gL1}108.6 & \cellcolor{gL1}107.2 \\
        & \textbf{pass@1-a (\%)} & \cellcolor{gL1}\textbf{86.87} & \cellcolor{gL1}84.38 & \cellcolor{gL1}85.00 & \cellcolor{gL3}44.53 & \cellcolor{gL1}84.84 & \cellcolor{gL1}83.75 \\
        \hline
        \multirow{18}{*}{LC} 
        & Init\_Var (3)
        & \cellcolor{gL2}\ \;\textbf{2.0} & \cellcolor{gL2}\ \;\textbf{2.0} & \cellcolor{gL2}\ \;1.6 & \cellcolor{gL3}\ \;1.0 & \cellcolor{gL2}\ \;\textbf{2.0} & \cellcolor{gL2}\ \;\textbf{2.0} \\
        & Repl\_Var (11)
        & \cellcolor{gL2}\ \;\textbf{7.6} & \cellcolor{gL2}\ \;6.2 & \cellcolor{gL2}\ \;7.0 & \cellcolor{gL3}\ \;5.0 & \cellcolor{gL2}\ \;6.6 & \cellcolor{gL2}\ \;6.4 \\
        & Upd\_Const (13)
        & \cellcolor{gL3}\ \;5.2 & \cellcolor{gL2}\ \;\textbf{8.6} & \cellcolor{gL2}\ \;7.0 & \ \;3.0 & \cellcolor{gL3}\ \;5.4 & \cellcolor{gL3}\ \;4.8 \\
        & Conv\_Type (23)
        & \cellcolor{gL2}17.2 & \cellcolor{gL2}16.4 & \cellcolor{gL1}\textbf{18.0} & \cellcolor{gL3}\ \;9.0 & \cellcolor{gL1}17.6 & \cellcolor{gL1}17.8 \\
        & Ins\_Call (22)
        & \cellcolor{gL2}15.0 & \cellcolor{gL2}13.8 & \cellcolor{gL2}\textbf{15.8} & \cellcolor{gL2}11.0 & \cellcolor{gL2}14.0 & \cellcolor{gL2}12.6 \\
        & Del\_Call (8)
        & \cellcolor{gL1}\ \;6.0 & \cellcolor{gL2}\ \;5.0 & \cellcolor{gL1}\ \;6.0 & \cellcolor{gL3}\ \;3.0 & \cellcolor{gL1}\ \;6.0 & \cellcolor{gL1}\ \;\textbf{7.2} \\
        & Upd\_Call (52)
        & \cellcolor{gL2}\textbf{32.8} & \cellcolor{gL2}\textbf{32.8} & \cellcolor{gL2}32.0 & \cellcolor{gL3}20.0 & \cellcolor{gL2}28.0 & \cellcolor{gL3}25.4 \\
        & Ins\_Import (16)
        & \cellcolor{gL3}\ \;\textbf{7.2} & \cellcolor{gL3}\ \;5.2 & \cellcolor{gL3}\ \;6.0 & \ \;1.0 & \cellcolor{gL3}\ \;4.4 & \ \;3.2 \\
        & Ins\_Return (5)
        & \cellcolor{gL3}\ \;2.2 & \cellcolor{gL2}\ \;\textbf{3.0} & \cellcolor{gL2}\ \;\textbf{3.0} & \cellcolor{gL3}\ \;2.0 & \cellcolor{gL2}\ \;\textbf{3.0} & \cellcolor{gL2}\ \;2.8 \\
        & Del\_Return (1)
        & \cellcolor{gL1}\ \;\textbf{1.0} & \cellcolor{gL1}\ \;0.8 & \cellcolor{gL1}\ \;\textbf{1.0} & \ \;0.0 & \ \;0.0 & \ \;0.0 \\
        & Upd\_Return (8)
        & \cellcolor{gL1}\ \;6.8 & \cellcolor{gL1}\ \;\textbf{7.8} & \cellcolor{gL1}\ \;6.0 & \cellcolor{gL1}\ \;6.0 & \cellcolor{gL1}\ \;6.0 & \cellcolor{gL1}\ \;6.8 \\
        & Ins\_Cond (31)
        & \cellcolor{gL2}19.2 & \cellcolor{gL2}20.4 & \cellcolor{gL2}\textbf{21.6} & \cellcolor{gL3}13.0 & \cellcolor{gL2}18.0 & \cellcolor{gL2}19.6 \\
        & Del\_Cond (13)
        & \cellcolor{gL1}\textbf{10.2} & \cellcolor{gL2}\ \;8.2 & \cellcolor{gL2}\ \;9.0 & \cellcolor{gL3}\ \;6.0 & \cellcolor{gL2}\ \;9.0 & \cellcolor{gL2}\ \;8.4 \\
        & Upd\_Cond (9)
        & \cellcolor{gL1}\ \;\textbf{9.0} & \cellcolor{gL1}\ \;8.2 & \cellcolor{gL1}\ \;\textbf{9.0} & \cellcolor{gL2}\ \;6.0 & \cellcolor{gL2}\ \;6.2 & \cellcolor{gL1}\ \;8.4 \\
        & Handle\_Ex (11)
        & \cellcolor{gL3}\ \;\textbf{4.2} & \cellcolor{gL3}\ \;4.0 & \cellcolor{gL3}\ \;4.0 & \cellcolor{gL3}\ \;3.0 & \cellcolor{gL3}\ \;\textbf{4.2} & \cellcolor{gL3}\ \;4.0 \\
        & Ins\_Func (1)
        & \cellcolor{gL1}\ \;\textbf{1.0} & \cellcolor{gL1}\ \;\textbf{1.0} & \cellcolor{gL1}\ \;\textbf{1.0} & \ \;0.0 & \cellcolor{gL1}\ \;\textbf{1.0} & \cellcolor{gL1}\ \;\textbf{1.0} \\
        \cline{2-8}
        & \textbf{All Types (227)}
        & \cellcolor{gL2}146.6 & \cellcolor{gL2}143.4 & \cellcolor{gL2}\textbf{148.0} & \cellcolor{gL3}\ \;89.0 & \cellcolor{gL2}131.4 & \cellcolor{gL2}130.4 \\ 
        & \textbf{pass@1-a (\%)} & \cellcolor{gL2}64.58 & \cellcolor{gL2}63.17 & \cellcolor{gL2}\textbf{65.20} & \cellcolor{gL3}39.21 & \cellcolor{gL2}57.89 & \cellcolor{gL2}57.44 \\
        \hline
        \multirow{18}{*}{RE}
        & Ins\_Ann (52)
        & \cellcolor{gL2}\textbf{31.8} & \cellcolor{gL2}28.4 & \cellcolor{gL2}28.6 & \cellcolor{gL3}13.0 & \cellcolor{gL2}28.6 & \cellcolor{gL2}29.0 \\
        & Upd\_Ann (3)
        & \cellcolor{gL3}\ \;1.0 & \cellcolor{gL3}\ \;1.0 & \cellcolor{gL3}\ \;1.0 & \ \;0.0 & \cellcolor{gL3}\ \;1.0 & \cellcolor{gL2}\ \;\textbf{2.0} \\
        & Del\_Ann (1)
        & \cellcolor{gL1}\ \;\textbf{1.0} & \cellcolor{gL1}\ \;\textbf{1.0} & \ \;0.0 & \ \;0.0 & \cellcolor{gL3}\ \;0.4 & \cellcolor{gL1}\ \;\textbf{1.0} \\
        & Ins\_Tmp\_Var (9)
        & \cellcolor{gL2}\ \;6.0 & \cellcolor{gL2}\ \;6.0 & \cellcolor{gL1}\ \;\textbf{7.0} & \cellcolor{gL3}\ \;3.0 & \cellcolor{gL2}\ \;6.2 & \cellcolor{gL2}\ \;4.8 \\
        & Del\_Tmp\_Var (3)
        & \cellcolor{gL2}\ \;\textbf{2.0} & \cellcolor{gL2}\ \;\textbf{2.0} & \cellcolor{gL2}\ \;\textbf{2.0} & \cellcolor{gL3}\ \;1.0 & \cellcolor{gL2}\ \;\textbf{2.0} & \cellcolor{gL2}\ \;\textbf{2.0} \\
        & Rename\_Param (24)
        & \cellcolor{gL1}19.0 & \cellcolor{gL1}18.6 & \cellcolor{gL1}19.0 & \cellcolor{gL3}\ \;8.0 & \cellcolor{gL2}17.6 & \cellcolor{gL1}\textbf{20.2} \\
        & Rename\_Var (20)
        & \cellcolor{gL1}16.0 & \cellcolor{gL1}15.0 & \cellcolor{gL1}16.0 & \cellcolor{gL2}11.0 & \cellcolor{gL2}14.6 & \cellcolor{gL1}\textbf{16.4} \\
        & Upd\_Api\_Ref (25)
        & \cellcolor{gL1}22.2 & \cellcolor{gL1}22.2 & \cellcolor{gL1}\textbf{22.8} & \cellcolor{gL2}14.0 & \cellcolor{gL1}20.2 & \cellcolor{gL1}21.4 \\
        & Ins\_Ret\_None (2)
        & \cellcolor{gL1}\ \;\textbf{2.0} & \cellcolor{gL1}\ \;\textbf{2.0} & \cellcolor{gL1}\ \;\textbf{2.0} & \cellcolor{gL2}\ \;1.0 & \cellcolor{gL1}\ \;\textbf{2.0} & \cellcolor{gL1}\ \;\textbf{2.0} \\
        & Refact\_Expr (17)
        & \cellcolor{gL1}13.8 & \cellcolor{gL1}\textbf{14.8} & \cellcolor{gL2}12.0 & \cellcolor{gL3}\ \;8.0 & \cellcolor{gL2}11.2 & \cellcolor{gL2}10.8 \\
        & Mov\_Expr (1)
        & \ \;0.0 & \ \;0.0 & \ \;0.0 & \ \;0.0 & \ \;0.0 & \ \;0.0 \\
        & Del\_Import (3)
        & \cellcolor{gL3}\ \;0.8 & \cellcolor{gL2}\ \;1.6 & \ \;0.4 & \cellcolor{gL3}\ \;1.0 & \cellcolor{gL2}\ \;1.8 & \cellcolor{gL2}\ \;\textbf{2.0} \\
        & Split\_Line (2)
        & \cellcolor{gL2}\ \;\textbf{1.0} & \cellcolor{gL2}\ \;\textbf{1.0} & \cellcolor{gL2}\ \;\textbf{1.0} & \cellcolor{gL2}\ \;\textbf{1.0} & \cellcolor{gL3}\ \;0.8 & \cellcolor{gL2}\ \;\textbf{1.0} \\
        & Inline\_Func (1)
        & \cellcolor{gL1}\ \;\textbf{1.0} & \cellcolor{gL1}\ \;\textbf{1.0} & \cellcolor{gL1}\ \;\textbf{1.0} & \cellcolor{gL1}\ \;\textbf{1.0} & \cellcolor{gL1}\ \;\textbf{1.0} & \ \;0.0 \\
        & Expd\_Func (3)
        & \cellcolor{gL2}\ \;1.8 & \cellcolor{gL2}\ \;\textbf{2.2} & \cellcolor{gL2}\ \;2.0 & \cellcolor{gL3}\ \;1.0 & \cellcolor{gL3}\ \;0.8 & \cellcolor{gL3}\ \;0.8 \\
        & Mov\_Func (2)
        & \cellcolor{gL2}\ \;\textbf{1.0} & \cellcolor{gL2}\ \;\textbf{1.0} & \cellcolor{gL2}\ \;\textbf{1.0} & \ \;0.0 & \ \;0.0 & \cellcolor{gL2}\ \;\textbf{1.0} \\
        \cline{2-8}
        & \textbf{All Types (168)}
        & \cellcolor{gL2}\textbf{120.4} & \cellcolor{gL2}117.8 & \cellcolor{gL2}115.8 & \cellcolor{gL3}\ \;63.0 & \cellcolor{gL2}108.2 & \cellcolor{gL2}114.4 \\
        & \textbf{pass@1-a (\%)} & \cellcolor{gL2}\textbf{71.67} & \cellcolor{gL2}70.12 & \cellcolor{gL2}68.93 & \cellcolor{gL3}37.50 & \cellcolor{gL2}64.40 & \cellcolor{gL2}68.10 \\
        \hline
        \multicolumn{2}{c|}{\textbf{All Categories (523)}}
        & \textbf{378.2} & 369.2 & 372.6 & 209.0 & 348.2 & 352.2 \\
        \multicolumn{2}{c|}{\textbf{pass@1-a (\%)}} & \textbf{72.31} & 70.59 & 71.24 & 39.96 & 66.58 & 67.34 \\
        \hline
    \end{tabular}
    \label{tab:res-type}
\end{table}

We further evaluate the \md{adaptation-level} performance of LLMs across each type under \md{the most informative} setting, i.e., \textbf{AReq} and \textit{Oracle}, \md{to explore LLMs' current strengths and limitations.}
The results are illustrated in Table~\ref{tab:res-type}, which is measured by pass@1-a. The darker color of the cell indicates a higher score. 
The best performing GPT-4o can solve 72.31\% adaptations on AdaptEval, while the worst Llama-3.3 can also achieve a pass@1-a about 40\%. Taking GPT-4o as an example, its pass@1-t in task-level evaluation is only 59.15\%. AdaptEval's fine-grained tests can reflect LLMs' performance more specifically, despite their failures in the overall task.

Based on our adaptation taxonomy, AdaptEval allows the evaluation of LLMs on different adaptation types. Specifically, LLMs obtain higher pass@1-a scores on \textit{Method Signature} than on \textit{Logic Customization}, with a 20.31\% gap on average.
We can observe that adaptations like function renaming are well-handled, with over 90\% of them are resolved by the top-tier LLMs. Most \textit{Logic Customization} adaptations have lower resolution rates, \textit{e.g.},  exception handling. The results indicate that LLMs excel at solving adaptations with straightforward solutions, while for adaptations require complex understanding and highly customized implementations, it is difficult for LLMs to carry out solutions that fully meet developers' needs.
Different LLMs have advantages in different adaptation types, \textit{e.g.}, GPT-4o, DeepSeek-V3, Gemini-2.0 and Codestral all achieve the best performance in more than 15 types. Although code LLMs perform comparably to general LLMs in \textit{Method Signature} and \textit{Refactoring}, they are less effective in logical adaptations. The possible reason may be their failures to fully understand the natural language instructions. 

\begin{center}
    \begin{tcolorbox}[colback=lgray,colframe=black,width=\linewidth,arc=1mm,boxrule=1pt,top=2pt,bottom=2pt,left=3pt,right=3pt]
        \textbf{Finding 2:} Based on AdaptEval's fine-grained evaluation, we find that LLMs perform best on \textit{Method Signature} and worst on \textit{Logic Customization}, with a gap of 20.31\%, as the latter requires complex understanding and implementation. 
    \end{tcolorbox}
\end{center}

\subsection{RQ3: Error Analysis}

\begin{table}[htbp]
    \scriptsize
    \centering
    \caption{The Error Distribution in LLMs' Adaptations.}
    \begin{tabular}{l|rrrr}
        \hline
        \multirow{2}{*}{\textbf{Model}} & \multicolumn{4}{c}{\textbf{Top Error Types}} \\
        \cline{2-5}
        & AssertionError & CompilationError & NameError & TypeError \\
        \hline
        GPT-4o & \textbf{226 (56.78\%) }& 69 (17.34\%)& 35 (\ \,8.79\%) & 28 (7.04\%) \\
        DS-V3 & \textbf{223 (54.93\%)} &  82 (20.20\%) &  48 (11.82\%) &  23 (5.67\%)\\
        Gemini-2.0 & \textbf{261 (59.59\%)} &  79 (18.04\%) &  40 (\ \,9.13\%) &  35 (7.99\%)\\
        Llama-3.3 &  105 (18.42\%) & \textbf{ 410 (71.93\%)} &  20 (\ \,3.51\%) &  10 (1.75\%)\\
        Qwen2.5 &  \textbf{304 (59.49\%)} & 80 (15.66\%) &  73 (14.29\%) &  24 (4.70\%) \\
        Codestral & \textbf{256 (53.78\%)} & 89 (18.70\%) &  68 (14.29\%) &  28 (5.88\%) \\
        \hline
    \end{tabular}
    \label{tab:res-errors}
\end{table}

\begin{figure}[htbp]
    \centering
    \includegraphics[width=\linewidth]{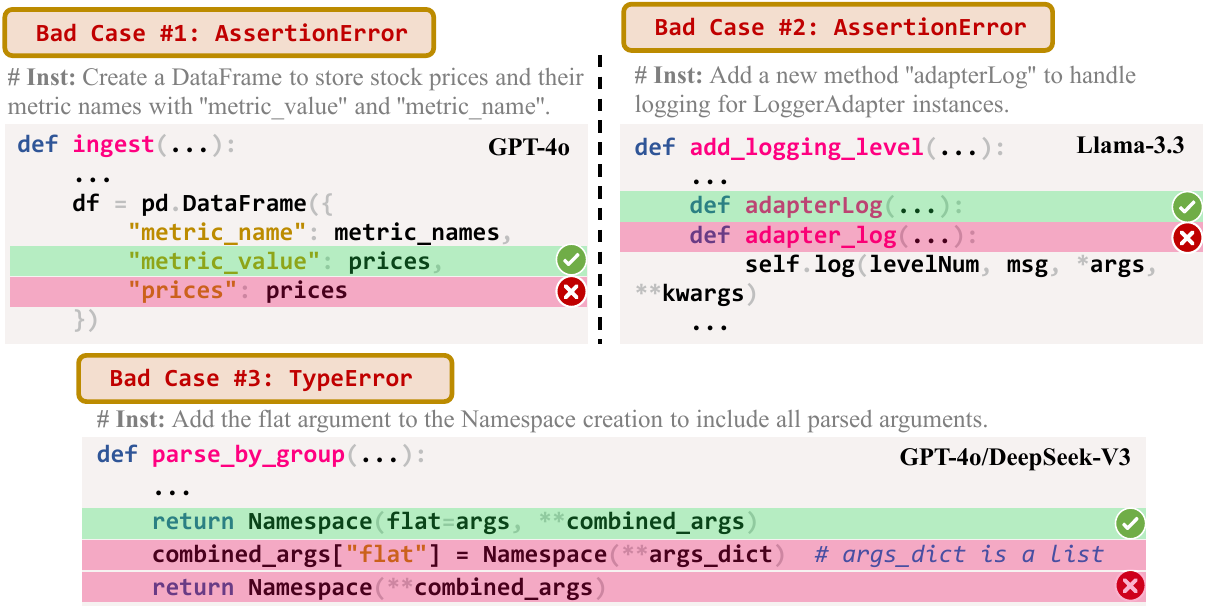}
    \caption{\md{Representative examples of LLMs' adaptation failures.}}
    \label{fig:bad-cases}
\end{figure}

This section further analyzes the errors in LLMs' adaptations under the \textbf{AReq} and \textit{Oracle} setting. Table~\ref{tab:res-errors} shows their error distributions. We identified 16 distinct types in total, but only retain the most frequent ones for clarity. 

\textit{AssertionError} is the dominant reason of failed adaptations. It indicates LLMs' misunderstanding or misimplementation of provided requirements. As shown in the first example of Figure~\ref{fig:bad-cases}, GPT-4o wrongly used the variable name ``\textit{prices}'' rather than the required ``\textit{metric\_value}'' to create the dataframe. In the second example, Llama-3.3 was instructed to add a new ``\textit{adapterLog}'' method but unexpectedly named it as ``\textit{adapt\_log}''. These failures highlight their reliance on the pre-training knowledge, \textit{e.g.}, the snake case naming style, rather than strictly following instructions.
For \textit{CompilationErrors}, the weakest LLM, Llama-3.3, often generates syntax-incorrect adaptations, whereas stronger LLMs rarely do. Such errors often arise during integration, such as incorrect method name.
The third most frequent type is \textit{NameError}, caused by undefined identifiers. We observe that code LLMs rise more errors in this type. Combined with results in Table~\ref{tab:res-type} (Ins\_Import and Del\_Import), this suggests their failures to include required libraries locally.
\textit{TypeErrors} arise from illegal operands or inconsistent arguments with method signature. In the last example, AdaptEval asked GPT-4o and DeepSeek-V3 to add a \textit{flat} argument. However, they used a \textit{list} rather than a \textit{dict} to create the ``\textit{Namespace}'' object, leading to a \textit{TypeError}.

\begin{center}
    \begin{tcolorbox}[colback=lgray,colframe=black,width=\linewidth,arc=1mm,boxrule=1pt,top=2pt,bottom=2pt,left=3pt,right=3pt]
        \textbf{Finding 3:} The most prevalent error raised by LLMs is \textit{AssertionError}, followed by \textit{CompilationError} and \textit{NameError}. We observe that LLMs may still hold on to their pre-training knowledge and violates our explicit instructions.
    \end{tcolorbox}
\end{center}

\subsection{RQ4: Potentials of Reasoning LLMs}

\begin{table}[htbp]
    \footnotesize
    \centering
    \caption{The Adaptation Performance of Reasoning LLMs.}
    \begin{tabular}{l|rr|rr|rr}
        \hline
        \multirow{2}{*}{\textbf{Model}} & \multicolumn{2}{c|}{\textbf{NoReq}} & \multicolumn{2}{c|}{\textbf{TReq}} & \multicolumn{2}{c}{\textbf{AReq}} \\ 
        & p@1-t & p@1-a & p@1-t & p@1-a & p@1-t & p@1-a \\
        \hline
        DS-R1 & \textbf{19.51} & \textbf{32.50} & 20.12 & 21.99 & 59.76 & \textbf{76.10} \\
        Claude-3.7  & 15.85 & 31.93 & \textbf{23.78} & \textbf{23.52} & \textbf{60.98} & 72.85 \\
        QwQ         & 15.24 & 23.52 & 18.90 & 19.50 & 56.71 & 71.89 \\
        \hline
        GPT-4o      & 12.19 & 26.58 & 20.73 & 19.92 & 59.15 & 72.31 \\ 
        Qwen2.5     &  9.63 & 21.80 & 16.83 & 16.14 & 48.66 & 66.58 \\
        \hline
    \end{tabular}
    \label{tab:reasoning-llms}
\end{table}

\begin{figure}[htbp]
    \centering
    \subfigure[Helpful reasoning (\textbf{NoReq})]{
        \includegraphics[width=\linewidth]{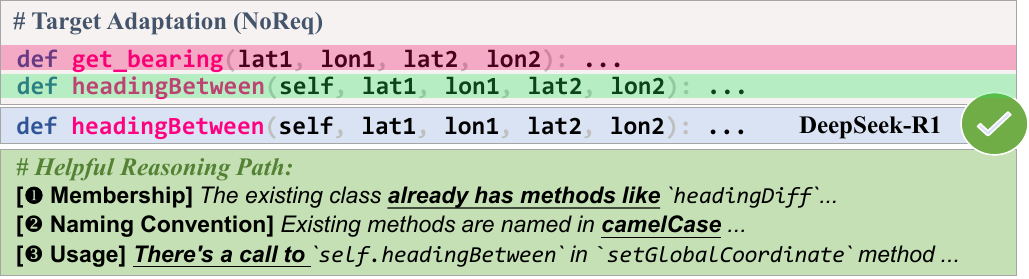}
        \label{fig:reasoning-example-a}
    }
    \subfigure[Harmful reasoning (\textbf{AReq})]{
        \includegraphics[width=\linewidth]{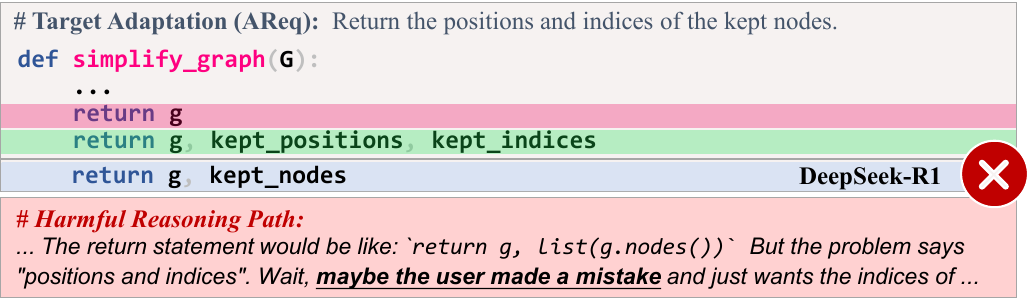}
        \label{fig:reasoning-example-b}
    }
    \caption{Adaptation examples of reasoning LLMs.}
    \label{fig:reasoning-examples}
\end{figure}

In this section, we investigate the potentials of reasoning LLMs on adaptation by conducting experiments in the \textbf{NoReq} setting and examining their reasoning paths. Table~\ref{tab:reasoning-llms} illustrates the adaptation performance of reasoning LLMs on AdaptEval. Compared with the best general and code LLMs, \textit{i.e.}, GPT-4o and Qwen2.5-Coder, reasoning LLMs achieve significantly higher accuracy in the \textbf{NoReq} setting, with up to 102\% higher pass@1-t. 
\md{We also observe that LLMs achieve higher pass@1-a score under \textbf{NoReq} than \textbf{TReq}, especially for DeepSeek-R1. It successfully solves 51 out of 114 \textit{Method Signature} adaptations that \textbf{TReq} fails to handle. However, in the 59 cases where \textbf{TReq} succeeds, 38 are \textit{Logic Customization}. Based on this,}
we further study the top types of adaptations that correctly completed by reasoning LLMs. As shown in Table~\ref{tab:reasoning-llms-type}, they excel at \textit{Method Signature} adaptations, including updating method type or renaming. The reason is that key information for these adaptations, \textit{e.g.}, usage or naming convention, is included in the context. An example of a helpful reasoning path of DeepSeek-R1 for adaptation is shown in Figure~\ref{fig:reasoning-example-a}. However, for adaptations like \textit{Logic Customization}, reasoning LLMs cannot precisely infer developers' intents merely from the context, resulting in their low accuracy. Our result highlights the effectiveness of reasoning LLMs in finding implicit contextual cues rather than speculating developers' requirements. 

Although reasoning LLMs in the \textbf{NoReq} setting approach their performance in \textbf{TReq}, they still lag significantly behind \textbf{AReq}. It underscores that while their reasoning capabilities can partially bridge the gap in requirements, further advancements are needed to fully align their reasoning process with developers' step-by-step adaptation strategies. 
It is also notable that instruction-tuned LLMs like GPT-4o achieves comparable performance with reasoning LLMs in \textbf{AReq}. This implies that their reasoning capability offers less assistance than in \textbf{NoReq}. 
To understand their limitations in following instructions, we \md{analyze the failed cases by manually inspecting their reasoning paths. We choose DeepSeek-R1 and GPT-4o for their best performance in the adaptation-level performance (pass@1-a). Among 523 adaptations, DeepSeek-R1 failed to perform 22 adaptations where the instruction-tuned GPT-4o succeeded. In 14 of them, DeepSeek-R1 expressed direct doubts to developers' instructions and refused to follow them.}  As shown in Figure 6(b), DeepSeek-R1 failed to update the return statement \md{by claiming that ``maybe the user made a mistake'' in its reasoning path. The remaining 8 cases include similar harmful self-reflection of its own correct reasoning instead of provided instructions.} However, GPT-4o correctly completed these adaptations by following the \md{instructions faithfully}. 
The result highlights that the improved reasoning ability makes LLMs more likely to reflect on their inputs and hence may sometimes harm their instruction-following ability. This also implies the importance of their further training on the requirement-driven adaptation data.

\begin{table}[htbp]
    \tiny
    \centering
    \caption{\md{Top Types of Adaptations Correctly Completed By Reasoning LLMs in the \emph{NoReq} setting.}}
    \begin{tabular}{l|cc|cc|cc}
        \hline
        \multirow{2}{*}{\textbf{Model}} & \multicolumn{2}{c|}{\textbf{Method Signature}} & \multicolumn{2}{c|}{\textbf{Logic Customization}} & \multicolumn{2}{c}{\textbf{Refactoring}} \\ 
        & \textbf{Type} & \multicolumn{1}{c|}{\textbf{p@1-a}} & \textbf{Type} & \multicolumn{1}{c|}{\textbf{p@1-a}} & \textbf{Type} & \multicolumn{1}{c}{\textbf{p@1-a}} \\
        \hline
        \multirow{4}{*}{DS-R1} 
        & Upd\_Type (11) & \cellcolor{gL1}81.82\% & Ins\_Import (16)  & \cellcolor{gL3}43.75\% & Del\_Import (3) & \cellcolor{gL2}66.66\% \\
        & Encap (13) & \cellcolor{gL2}61.54\% & Repl\_Var (11) & \cellcolor{gL3}36.36\% & Upd\_Api\_Ref (25)  & \cellcolor{gL2}64.00\% \\
        & Rnm\_Func (70) & \cellcolor{gL2}60.00\% & Init\_Var (3) & \cellcolor{gL3}33.33\% & Del\_Tmp\_Var (3) & \cellcolor{gL3}33.33\% \\
        \cline{2-7}
        & \textbf{Average}  & 50.78\% & \textbf{Average} & 25.11\% & \textbf{Average}  & 28.57\% \\
        \hline
        \multirow{4}{*}{Claude} 
        & Upd\_Type (11) & \cellcolor{gL1}90.91\% & Handle\_Ex (11)  & \cellcolor{gL3}45.45\% & Upd\_Api\_Ref (25)  & \cellcolor{gL2}68.00\% \\
        & Rnm\_Func (70) & \cellcolor{gL2}62.86\% & Ins\_Return (5) & \cellcolor{gL3}40.00\% & Del\_Import (3) & \cellcolor{gL2}66.67\% \\
        & Encap (13) & \cellcolor{gL2}53.85\% & Ins\_Import (16) & \cellcolor{gL3}37.50\%  & Ins\_Ret\_None (2)  & \cellcolor{gL2}50.00\% \\
        \cline{2-7}
        & \textbf{Average}  & 51.56\% & \textbf{Average} & 24.67\% & \textbf{Average}  & 24.40\% \\
        \hline
        \multirow{4}{*}{QwQ}
        & Upd\_Type (11) & \cellcolor{gL2}54.55\% & Handle\_Ex (11)  & \cellcolor{gL3}45.45\% & Del\_Import (3)  & \cellcolor{gL2}66.66\% \\
        & Encap (13) & \cellcolor{gL2}53.85\% & Ins\_Return (5) & \cellcolor{gL3}40.00\% & Upd\_Api\_Ref (25) & \cellcolor{gL2}64.00\% \\
        & Rnm\_Func (70) & \cellcolor{gL3}44.29\% & Init\_Var (3)  & \cellcolor{gL3}33.33\% & Ins\_Ret\_None (2) & \cellcolor{gL2}50.00\% \\
        \cline{2-7}
        & \textbf{Average}  & 37.50\% & \textbf{Average} & 18.06\% & \textbf{Average} & 20.24\% \\
        \hline
    \end{tabular}
    \label{tab:reasoning-llms-type}
\end{table}

\begin{center}
    \begin{tcolorbox}[colback=lgray,colframe=black,width=\linewidth,arc=1mm,boxrule=1pt,top=2pt,bottom=2pt,left=3pt,right=3pt]
        \textbf{Finding 4:} Reasoning LLMs significantly outperform instruction-tuned LLMs in adaptation when no requirements are provided. However, their reasoning process still need more alignments to developers' actual adaptations.
    \end{tcolorbox}
\end{center}

\section{Discussion}
\subsection{Data Leakage Issue}

\begin{figure}[htbp]
    \centering
    \includegraphics[width=\linewidth]{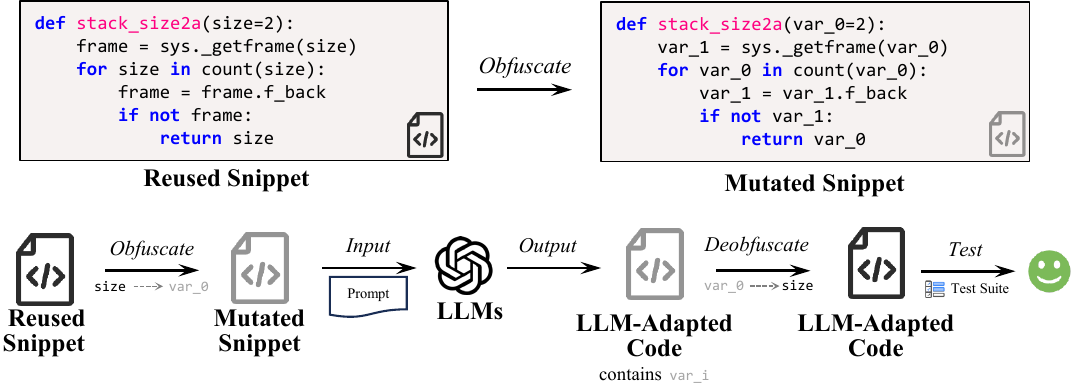}
    \caption{\md{The mutation-based approach to detecting and mitigating data leakage.}}
    \label{fig:mutation}
\end{figure}

\begin{table}[ht]
    \scriptsize
    \centering
    \caption{The Extent of Data Leakage In AdaptEval.}
    \label{tab:data-leakage}
    \begin{tabular}{cccc}
        \hline
        \textbf{Model} & \textbf{Leak Ratio (\%)} & \textbf{$\Delta$Pass@1-t} & \textbf{$\Delta$Pass@1-a} \\
        \hline
        GPT-4o        & 1.22 & -2.45 (-4.14\%) & -2.14 (-2.96\%) \\
        DeepSeek-V3   & 1.22 & -3.04 (-5.30\%) &  -2.71 (-3.84\%) \\
        Gemini-2.0    & 1.22  & \textbf{-4.02 (-7.36\%)}  & \textbf{-4.51 (-6.33\%)}\\
        Llama-3.3     & 0.61 & -1.83 (-5.36\%) & -1.53 (-3.83\%) \\
        Qwen2.5-Coder & 0.61 & -1.71 (-3.51\%) & -1.76 (-2.64\%)\\
        Codestral     & 0.61  & -2.20 (-4.32\%) & -0.99 (-1.47\%)\\
        \hline
        DeepSeek-R1 & 1.22 & -0.62 (-1.04\%) & -2.10 (-2.76\%)\\
        Claude-3.7  & 1.22  & -0.61 (-1.00\%) &-0.57 (-0.78\%) \\
        QwQ-32B         & 1.22 & +0.61 (+1.08\%) & -2.10 (-2.92\%)\\
        \hline
    \end{tabular}  
\end{table}

\md{As current LLMs may include various data sources including GitHub in their training corpus, evaluating them on the leaked data may impact the effectiveness of benchmarks. It is essential to measure the extent of data leakage of AdaptEval on our selected LLMs and provide leakage mitigation approaches for its future use. In our study, we adopt CDD~\cite{Dong2024} and a mutation-based evaluation~\cite{Kong2025,Zhang2025Unseen} to detect data leakage. Specifically, CDD assumes that LLMs with the data leakage issue are more prone to generate outputs that resemble their training data. Following their assumption, we check whether the sampled outputs of our used LLMs are highly similar to the possibly leaked data. Regarding the mutation-based approach, it first modifies the original code input to LLMs before adaptation, i.e., renaming all the identifiers, and compare LLMs' performance in adapting the mutated code with the original one. If they use their memorization of the leaked data, their performance on the unseen data, i.e., the mutated code, will be degraded. Otherwise, the extent of performance decrease should be low because they perform the task with the same reasoning ability. Our implemented mutation-based approach is depicted in Figure~\ref{fig:mutation}. The evaluation results of AdaptEval's data leakage issue are shown in Table~\ref{tab:data-leakage}. Compared to the benchmarks suffered from data leakage, i.e., a 41.17\% Leak Ratio on HumanEval and a 52.27\% on Defect4J (both evaluated with GPT-3.5), AdaptEval only exhibits a leak ratio of less than 2\% and an up to 7.36\% degradation in the pass@1 score, suggesting a low risk of data leakage on all LLMs used. Besides, our mutation-based approach could also mitigate the data leakage risk in future evaluation. Even if the ground truth code is learned by LLMs, we can use the mutated code to conduct evaluation on AdaptEval.}  

\subsection{Implications}
\textbf{We pose a new benchmark to advance research and practice in \md{code snippet} reuse.} While \md{code snippet} reuse is ubiquitous in development, the adaptation process remains a significant bottleneck. Unlike existing code generation benchmarks, AdaptEval provides a focused evaluation framework for assessing LLMs' adaptations ability. It opens new directions to improve LLMs for reuse-specific tasks and calls for tools to streamline adaptation and reduce human efforts.

\textbf{\md{LLMs' pre-training knowledge may hinders their ability to understand and adhere to user instructions.}} 
The adaptation-level annotations in AdaptEval \md{enable a fine-grained analysis of LLMs' instruction following ability}. Our findings show that although LLMs significantly benefit from \md{specific and actionable guidance, they may still fail to follow the explicit instructions. This suggests LLMs' weakness in handling the conflict between their internal pre-training knowledge and external instructions. For researchers, our study highlights the need to investigate methods for better coordination between LLMs' instruction following abilities and pre-training knowledge to enhance their performance.}

\textbf{More task-specific and context-aware techniques are expected to facilitate the resolution of the adaptation task.} Though being prompted with specific instructions, current LLMs may still struggle with complex adaptations, \textit{e.g.}, \textit{Logic Customization}) due to their lack of pre-existing knowledge, accurate reasoning and context understanding. To this end, future studies could consider specialized training on sub-tasks such as dependency handling, or retrieval-augmented techniques to utilize associated contextual information, including SO discussions and development context.

\section{Threats to Validity}
\label{sec:threats}
\textbf{Threats in data collection.} \md{AdaptEval only includes the scenario of adapting code snippets from Stack Overflow. Our evaluation results may not be generalized to other adaptation scenarios, as well as future software development process as LLMs become better at code generation.}
\md{Although our SO post filtration considers the data coverage, it may miss valuable adaptations on posts with negative scores, e.g., corrective changes to address quality issues.}
To filter spurious code reuse, we combine link identification with clone detection techniques. However, this may ignore cases without references or with substantial adaptations. 
\textbf{Threats in data annotation.} Human annotation of adaptation types and descriptions may introduce subjective bias. To alleviate this issue, we employ a rigorous annotation process by referring the actual commit history and ensuring final annotations are approved by all annotators.
\textbf{Threats in benchmark scale.} Another potential threat is the benchmark scale. Considering data leakage and the authenticity of code reuse, we try our best to include all available cases that satisfy our criteria. 
However, we only focus on \md{adaptations in a single programming language and under several controlled scenarios. Our findings may not generalize to adaptations in other languages and scenarios. Hence, we plan to extend our benchmark to support multiple languages and more scenarios in the future work.}
\textbf{Threats in evaluation.} The manual creation of prompts and tests may bias our evaluation results. To mitigate this, we follow the best practices~\cite{Zhao2023} for prompt design. For test cases, we conduct a rigorous checking process and plan to release them to the community for validation. 
\md{The adaptation-level evaluation are only conducted in the ideal setting for investigation current LLMs' boundaries, while our conclusion may not be held in other scenarios.}
\md{Though we adopt a data leakage mitigation strategy for AdaptEval, there will still be chance that LLMs learn our benchmark in the future. We will continue to maintain and update our benchmark for the community.}

\section{Conclusion}
This paper proposes AdaptEval benchmark to evaluate LLMs' capabilities in code snippet adaptation. It is equipped with practical context, multi-granular annotations and fine-grained evaluation. Based on AdaptEval, we conduct an empirical study on six instruction-tuned LLMs and especially three reasoning LLMs. Results demonstrate that LLMs perform better with actionable instructions than task-level intentions. They achieve lower performance in adaptations that require complex understanding or implementation. For reasoning LLMs, they excel at inferring implicit contextual cues but their reasoning process still deviates from developers' actual adaptation strategies. \md{Our benchmark supports further exploration of applying LLMs in code reuse and adaptation tasks.} 

\section*{Acknowledgments}
We gratefully acknowledge the support from  the National Key Research and Development Program of China (Grant No.2023YFB4503802) and the National Natural Science Foundation of China (Grant No.62302515, No.62172426, and No.62332005).

\bibliographystyle{ieeetr}
\bibliography{ref}

@inproceedings{Du2024,
author = {Du, Xueying and Liu, Mingwei and Wang, Kaixin and Wang, Hanlin and Liu, Junwei and Chen, Yixuan and Feng, Jiayi and Sha, Chaofeng and Peng, Xin and Lou, Yiling},
title = {Evaluating Large Language Models in Class-Level Code Generation},
year = {2024},
isbn = {9798400702174},
publisher = {Association for Computing Machinery},
address = {New York, NY, USA},
url = {https://doi.org/10.1145/3597503.3639219},
doi = {10.1145/3597503.3639219},
booktitle = {Proceedings of the IEEE/ACM 46th International Conference on Software Engineering},
articleno = {81},
numpages = {13},
location = {, Lisbon, Portugal, },
series = {ICSE '24}
}

@inproceedings{Zhang2019,
	series = {{ICSE} '19},
	title = {Analyzing and {Supporting} {Adaptation} of {Online} {Code} {Examples}},
	url = {https://doi.org/10.1109/ICSE.2019.00046},
	doi = {10.1109/ICSE.2019.00046},
	booktitle = {Proceedings of the 41st {International} {Conference} on {Software} {Engineering}},
	publisher = {IEEE Press},
	author = {Zhang, Tianyi and Yang, Di and Lopes, Crista and Kim, Miryung},
	year = {2019},
	note = {event-place: Montreal, Quebec, Canada},
	pages = {316--327}
}

@inproceedings{Falleri2014,
	address = {Vasteras Sweden},
	title = {Fine-grained and accurate source code differencing},
	isbn = {978-1-4503-3013-8},
	url = {https://dl.acm.org/doi/10.1145/2642937.2642982},
	doi = {10.1145/2642937.2642982},
	language = {en},
	urldate = {2022-04-14},
	booktitle = {Proceedings of the 29th {ACM}/{IEEE} international conference on {Automated} software engineering},
	publisher = {ACM},
	author = {Falleri, Jean-Rémy and Morandat, Floréal and Blanc, Xavier and Martinez, Matias and Monperrus, Martin},
	month = sep,
	year = {2014},
	pages = {313--324},
}

@inproceedings{Peng2024,
    title = "{H}uman{E}val-{XL}: A Multilingual Code Generation Benchmark for Cross-lingual Natural Language Generalization",
    author = "Peng, Qiwei  and
      Chai, Yekun  and
      Li, Xuhong",
    editor = "Calzolari, Nicoletta  and
      Kan, Min-Yen  and
      Hoste, Veronique  and
      Lenci, Alessandro  and
      Sakti, Sakriani  and
      Xue, Nianwen",
    booktitle = "Proceedings of the 2024 Joint International Conference on Computational Linguistics, Language Resources and Evaluation (LREC-COLING 2024)",
    month = may,
    year = "2024",
    address = "Torino, Italia",
    publisher = "ELRA and ICCL",
    url = "https://aclanthology.org/2024.lrec-main.735",
    pages = "8383--8394",
}

@inproceedings{Yu2024,
author = {Yu, Hao and Shen, Bo and Ran, Dezhi and Zhang, Jiaxin and Zhang, Qi and Ma, Yuchi and Liang, Guangtai and Li, Ying and Wang, Qianxiang and Xie, Tao},
title = {CoderEval: A Benchmark of Pragmatic Code Generation with Generative Pre-trained Models},
year = {2024},
isbn = {9798400702174},
publisher = {Association for Computing Machinery},
address = {New York, NY, USA},
url = {https://doi.org/10.1145/3597503.3623316},
doi = {10.1145/3597503.3623316},
booktitle = {Proceedings of the IEEE/ACM 46th International Conference on Software Engineering},
articleno = {37},
numpages = {12},
location = {Lisbon, Portugal},
series = {ICSE '24}
}

@misc{Du2024_Mercury,
      title={Mercury: A Code Efficiency Benchmark for Code Large Language Models}, 
      author={Mingzhe Du and Anh Tuan Luu and Bin Ji and Qian Liu and See-Kiong Ng},
      year={2024},
      eprint={2402.07844},
      archivePrefix={arXiv},
      primaryClass={cs.SE},
      url={https://arxiv.org/abs/2402.07844}, 
}

@misc{Huang2024,
      title={EffiBench: Benchmarking the Efficiency of Automatically Generated Code}, 
      author={Dong Huang and Yuhao Qing and Weiyi Shang and Heming Cui and Jie M. Zhang},
      year={2024},
      eprint={2402.02037},
      archivePrefix={arXiv},
      primaryClass={cs.SE},
      url={https://arxiv.org/abs/2402.02037}, 
}

@misc{Zeng2024,
      title={CoderUJB: An Executable and Unified Java Benchmark for Practical Programming Scenarios}, 
      author={Zhengran Zeng and Yidong Wang and Rui Xie and Wei Ye and Shikun Zhang},
      year={2024},
      eprint={2403.19287},
      archivePrefix={arXiv},
      primaryClass={cs.SE},
      url={https://arxiv.org/abs/2403.19287}, 
}

@misc{Li2024,
      title={DevEval: A Manually-Annotated Code Generation Benchmark Aligned with Real-World Code Repositories}, 
      author={Jia Li and Ge Li and Yunfei Zhao and Yongmin Li and Huanyu Liu and Hao Zhu and Lecheng Wang and Kaibo Liu and Zheng Fang and Lanshen Wang and Jiazheng Ding and Xuanming Zhang and Yuqi Zhu and Yihong Dong and Zhi Jin and Binhua Li and Fei Huang and Yongbin Li},
      year={2024},
      eprint={2405.19856},
      archivePrefix={arXiv},
      primaryClass={cs.CL},
      url={https://arxiv.org/abs/2405.19856}, 
}

@misc{Xie2024,
      title={CodeBenchGen: Creating Scalable Execution-based Code Generation Benchmarks}, 
      author={Yiqing Xie and Alex Xie and Divyanshu Sheth and Pengfei Liu and Daniel Fried and Carolyn Rose},
      year={2024},
      eprint={2404.00566},
      archivePrefix={arXiv},
      primaryClass={cs.SE},
      url={https://arxiv.org/abs/2404.00566}, 
}

@misc{Deng2024,
      title={R2C2-Coder: Enhancing and Benchmarking Real-world Repository-level Code Completion Abilities of Code Large Language Models}, 
      author={Ken Deng and Jiaheng Liu and He Zhu and Congnan Liu and Jingxin Li and Jiakai Wang and Peng Zhao and Chenchen Zhang and Yanan Wu and Xueqiao Yin and Yuanxing Zhang and Wenbo Su and Bangyu Xiang and Tiezheng Ge and Bo Zheng},
      year={2024},
      eprint={2406.01359},
      archivePrefix={arXiv},
      primaryClass={cs.CL},
      url={https://arxiv.org/abs/2406.01359}, 
}

@inproceedings{Ding2023,
 author = {Ding, Yangruibo and Wang, Zijian and Ahmad, Wasi and Ding, Hantian and Tan, Ming and Jain, Nihal and Ramanathan, Murali Krishna and Nallapati, Ramesh and Bhatia, Parminder and Roth, Dan and Xiang, Bing},
 booktitle = {Advances in Neural Information Processing Systems},
 editor = {A. Oh and T. Naumann and A. Globerson and K. Saenko and M. Hardt and S. Levine},
 pages = {46701--46723},
 publisher = {Curran Associates, Inc.},
 title = {CrossCodeEval: A Diverse and Multilingual Benchmark for Cross-File Code Completion},
 url = {https://proceedings.neurips.cc/paper_files/paper/2023/file/920f2dced7d32ab2ba2f1970bc306af6-Paper-Datasets_and_Benchmarks.pdf},
 volume = {36},
 year = {2023}
}

@INPROCEEDINGS{Niu2023,
  author={Niu, Changan and Li, Chuanyi and Ng, Vincent and Luo, Bin},
  booktitle={2023 IEEE/ACM 45th International Conference on Software Engineering (ICSE)}, 
  title={CrossCodeBench: Benchmarking Cross-Task Generalization of Source Code Models}, 
  year={2023},
  volume={},
  number={},
  pages={537-549},
  doi={10.1109/ICSE48619.2023.00055}}

@misc{Liu2023,
      title={RepoBench: Benchmarking Repository-Level Code Auto-Completion Systems}, 
      author={Tianyang Liu and Canwen Xu and Julian McAuley},
      year={2023},
      eprint={2306.03091},
      archivePrefix={arXiv},
      primaryClass={cs.CL},
      url={https://arxiv.org/abs/2306.03091}, 
}

@misc{Khan2023,
      title={xCodeEval: A Large Scale Multilingual Multitask Benchmark for Code Understanding, Generation, Translation and Retrieval}, 
      author={Mohammad Abdullah Matin Khan and M Saiful Bari and Xuan Long Do and Weishi Wang and Md Rizwan Parvez and Shafiq Joty},
      year={2023},
      eprint={2303.03004},
      archivePrefix={arXiv},
      primaryClass={cs.CL},
      url={https://arxiv.org/abs/2303.03004}, 
}

@misc{Fu2024,
      title={CodeApex: A Bilingual Programming Evaluation Benchmark for Large Language Models}, 
      author={Lingyue Fu and Huacan Chai and Shuang Luo and Kounianhua Du and Weiming Zhang and Longteng Fan and Jiayi Lei and Renting Rui and Jianghao Lin and Yuchen Fang and Yifan Liu and Jingkuan Wang and Siyuan Qi and Kangning Zhang and Weinan Zhang and Yong Yu},
      year={2024},
      eprint={2309.01940},
      archivePrefix={arXiv},
      primaryClass={cs.CL},
      url={https://arxiv.org/abs/2309.01940}, 
}

@misc{Yan2024,
      title={CodeScope: An Execution-based Multilingual Multitask Multidimensional Benchmark for Evaluating LLMs on Code Understanding and Generation}, 
      author={Weixiang Yan and Haitian Liu and Yunkun Wang and Yunzhe Li and Qian Chen and Wen Wang and Tingyu Lin and Weishan Zhao and Li Zhu and Hari Sundaram and Shuiguang Deng},
      year={2024},
      eprint={2311.08588},
      archivePrefix={arXiv},
      primaryClass={cs.CL},
      url={https://arxiv.org/abs/2311.08588}, 
}

@misc{Guo2024,
      title={CodeEditorBench: Evaluating Code Editing Capability of Large Language Models}, 
      author={Jiawei Guo and Ziming Li and Xueling Liu and Kaijing Ma and Tianyu Zheng and Zhouliang Yu and Ding Pan and Yizhi LI and Ruibo Liu and Yue Wang and Shuyue Guo and Xingwei Qu and Xiang Yue and Ge Zhang and Wenhu Chen and Jie Fu},
      year={2024},
      eprint={2404.03543},
      archivePrefix={arXiv},
      primaryClass={cs.SE},
      url={https://arxiv.org/abs/2404.03543}, 
}

@misc{Deshpande2024,
      title={Class-Level Code Generation from Natural Language Using Iterative, Tool-Enhanced Reasoning over Repository}, 
      author={Ajinkya Deshpande and Anmol Agarwal and Shashank Shet and Arun Iyer and Aditya Kanade and Ramakrishna Bairi and Suresh Parthasarathy},
      year={2024},
      eprint={2405.01573},
      archivePrefix={arXiv},
      primaryClass={cs.SE},
      url={https://arxiv.org/abs/2405.01573}, 
}

@misc{Jimenez2024,
      title={SWE-bench: Can Language Models Resolve Real-World GitHub Issues?}, 
      author={Carlos E. Jimenez and John Yang and Alexander Wettig and Shunyu Yao and Kexin Pei and Ofir Press and Karthik Narasimhan},
      year={2024},
      eprint={2310.06770},
      archivePrefix={arXiv},
      primaryClass={cs.CL},
      url={https://arxiv.org/abs/2310.06770}, 
}

@misc{Tian2024,
      title={DebugBench: Evaluating Debugging Capability of Large Language Models}, 
      author={Runchu Tian and Yining Ye and Yujia Qin and Xin Cong and Yankai Lin and Yinxu Pan and Yesai Wu and Haotian Hui and Weichuan Liu and Zhiyuan Liu and Maosong Sun},
      year={2024},
      eprint={2401.04621},
      archivePrefix={arXiv},
      primaryClass={cs.SE},
      url={https://arxiv.org/abs/2401.04621}, 
}

@misc{Li2024DevBench,
      title={DevBench: A Comprehensive Benchmark for Software Development}, 
      author={Bowen Li and Wenhan Wu and Ziwei Tang and Lin Shi and John Yang and Jinyang Li and Shunyu Yao and Chen Qian and Binyuan Hui and Qicheng Zhang and Zhiyin Yu and He Du and Ping Yang and Dahua Lin and Chao Peng and Kai Chen},
      year={2024},
      eprint={2403.08604},
      archivePrefix={arXiv},
      primaryClass={cs.CL},
      url={https://arxiv.org/abs/2403.08604}, 
}

@ARTICLE{Zhang2024,
  author={Zhang, Tanghaoran and Lu, Yao and Yu, Yue and Mao, Xinjun and Zhang, Yang and Zhao, Yuxin},
  journal={IEEE Transactions on Software Engineering}, 
  title={How do Developers Adapt Code Snippets to Their Contexts? An Empirical Study of Context-Based Code Snippet Adaptations}, 
  year={2024},
  volume={},
  number={},
  pages={1-20},
  doi={10.1109/TSE.2024.3395519}}

@inproceedings{Terragni2021,
	address = {Melbourne, Australia},
	title = {{APIzation}: {Generating} {Reusable} {APIs} from {StackOverflow} {Code} {Snippets}},
	isbn = {978-1-66540-337-5},
	shorttitle = {{APIzation}},
	url = {https://ieeexplore.ieee.org/document/9678576/},
	doi = {10.1109/ASE51524.2021.9678576},
	language = {en},
	urldate = {2023-12-15},
	booktitle = {2021 36th {IEEE}/{ACM} {International} {Conference} on {Automated} {Software} {Engineering} ({ASE})},
	publisher = {IEEE},
	author = {Terragni, Valerio and Salza, Pasquale},
	month = nov,
	year = {2021},
	pages = {542--554},
}

@article{Wu2019,
	title = {How do developers utilize source code from stack overflow?},
	volume = {24},
	issn = {1382-3256, 1573-7616},
	url = {http://link.springer.com/10.1007/s10664-018-9634-5},
	doi = {10.1007/s10664-018-9634-5},
	language = {en},
	number = {2},
	urldate = {2022-06-06},
	journal = {Empirical Software Engineering},
	author = {Wu, Yuhao and Wang, Shaowei and Bezemer, Cor-Paul and Inoue, Katsuro},
	month = apr,
	year = {2019},
	pages = {637--673},
}

@inproceedings{Mondal2019,
	address = {Cleveland, OH, USA},
	title = {Investigating {Context} {Adaptation} {Bugs} in {Code} {Clones}},
	isbn = {978-1-72813-094-1},
	url = {https://ieeexplore.ieee.org/document/8919065/},
	doi = {10.1109/ICSME.2019.00026},
	language = {en},
	urldate = {2022-09-15},
	booktitle = {2019 {IEEE} {International} {Conference} on {Software} {Maintenance} and {Evolution} ({ICSME})},
	publisher = {IEEE},
	author = {Mondal, Manishankar and Roy, Banani and Roy, Chanchal K. and Schneider, Kevin A.},
	month = sep,
	year = {2019},
	pages = {157--168},
}

@inproceedings{Huang2022,
	address = {New York, NY, USA},
	series = {{ICPC} '22},
	title = {Towards {Exploring} the {Code} {Reuse} from {Stack} {Overflow} during {Software} {Development}},
	isbn = {978-1-4503-9298-3},
	url = {https://doi.org/10.1145/3524610.3527923},
	doi = {10.1145/3524610.3527923},
	booktitle = {Proceedings of the 30th {IEEE}/{ACM} {International} {Conference} on {Program} {Comprehension}},
	publisher = {Association for Computing Machinery},
	author = {Huang, Yuan and Xu, Furen and Zhou, Haojie and Chen, Xiangping and Zhou, Xiaocong and Wang, Tong},
	year = {2022},
	note = {event-place: Virtual Event},
	pages = {548--559},
}

@inproceedings{Dabic2021,
  author    = {Ozren Dabic and Emad Aghajani and Gabriele Bavota},
  title     = {Sampling Projects in GitHub for {MSR} Studies},
  booktitle = {18th {IEEE/ACM} International Conference on Mining Software Repositories,
               {MSR} 2021},
  pages     = {560--564},
  publisher = {{IEEE}},
  year      = {2021}
}

@inproceedings{Radford2018,
  title={Improving Language Understanding by Generative Pre-Training},
  author={Alec Radford and Karthik Narasimhan},
  year={2018},
  url={https://api.semanticscholar.org/CorpusID:49313245}
}

@inproceedings{Radford2019,
  title={Language Models are Unsupervised Multitask Learners},
  author={Alec Radford and Jeff Wu and Rewon Child and David Luan and Dario Amodei and Ilya Sutskever},
  year={2019},
  url={https://api.semanticscholar.org/CorpusID:160025533}
}

@misc{Brown2020,
	title = {Language {Models} are {Few}-{Shot} {Learners}},
	url = {http://arxiv.org/abs/2005.14165},
	language = {en},
	urldate = {2023-02-20},
	publisher = {arXiv},
	author = {Brown, Tom B. and Mann, Benjamin and Ryder, Nick and Subbiah, Melanie and Kaplan, Jared and Dhariwal, Prafulla and Neelakantan, Arvind and Shyam, Pranav and Sastry, Girish and Askell, Amanda and others},
	month = jul,
	year = {2020},
	note = {arXiv:2005.14165 [cs]},
}

@online{OpenAI2023,
  author={OpenAI},
  title={gpt-3.5-turbo},
  year=2023,
  url={https://platform.openai.com/docs/models/gpt-3-5-turbo},
}

@online{OpenAI2024,
  author={OpenAI},
  title={gpt-4o},
  year=2024,
  url={https://platform.openai.com/docs/models/gpt-4o},
}

@online{Anthropic2025,
  author={Anthropic},
  title={Claude-3.7-Sonnet},
  year=2025,
  url={https://www.anthropic.com/news/claude-3-7-sonnet},
}

@online{Google2025,
  author={Google-DeepMind},
  title={Gemini-2.0-Flash},
  year=2025,
  url={https://deepmind.google/models/gemini},
}

@online{MistralAI2024,
  author={MistralAI},
  title={Codestral},
  year=2024,
  url={https://docs.mistral.ai/getting-started/models},
}

@online{Meta2024,
  author={Meta},
  title = {Meta Llama 3},
  year=2024,
  url= {https://llama.meta.com/},
}

@article{Chen2021,
  title={Evaluating Large Language Models Trained on Code},
  author={Mark Chen and Jerry Tworek and Heewoo Jun and Qiming Yuan and Henrique Ponde de Oliveira Pinto and Jared Kaplan and Harri Edwards and Yuri Burda and Nicholas Joseph and Greg Brockman and others},
  year={2021},
  eprint={2107.03374},
  archivePrefix={arXiv},
  primaryClass={cs.LG}
}

@misc{Deepseek-ai_2024b,
	title = {{DeepSeek}-{Coder}-{V2}: {Breaking} the {Barrier} of {Closed}-{Source} {Models} in {Code} {Intelligence}},
	shorttitle = {{DeepSeek}-{Coder}-{V2}},
	url = {http://arxiv.org/abs/2406.11931},
	language = {en},
	urldate = {2024-08-21},
	publisher = {arXiv},
	author = {DeepSeek-AI and Zhu, Qihao and Guo, Daya and Shao, Zhihong and Yang, Dejian and Wang, Peiyi and Xu, Runxin and Wu, Y. and Li, Yukun and Gao, Huazuo and others},
	month = jun,
	year = {2024},
	note = {arXiv:2406.11931 [cs]},
}

@inproceedings{Yang2017,
	title = {Stack {Overflow} in {Github}: {Any} {Snippets} {There}?},
	doi = {10.1109/MSR.2017.13},
	booktitle = {2017 {IEEE}/{ACM} 14th {International} {Conference} on {Mining} {Software} {Repositories} ({MSR})},
	author = {Yang, Di and Martins, Pedro and Saini, Vaibhav and Lopes, Cristina},
	year = {2017},
	pages = {280--290},
}

@article{Baltes2019,
	title = {Usage and attribution of {Stack} {Overflow} code snippets in {GitHub} projects},
	volume = {24},
	issn = {1382-3256, 1573-7616},
	url = {http://link.springer.com/10.1007/s10664-018-9650-5},
	doi = {10.1007/s10664-018-9650-5},
	language = {en},
	number = {3},
	urldate = {2022-02-12},
	journal = {Empirical Software Engineering},
	author = {Baltes, Sebastian and Diehl, Stephan},
	month = jun,
	year = {2019},
	pages = {1259--1295}
}

@inproceedings{Cottrell2008,
	address = {Leipzig, Germany},
	title = {Jigsaw: a tool for the small-scale reuse of source code},
	isbn = {978-1-60558-079-1},
	shorttitle = {Jigsaw},
	url = {http://portal.acm.org/citation.cfm?doid=1370175.1370194},
	doi = {10.1145/1370175.1370194},
	language = {en},
	urldate = {2022-06-07},
	booktitle = {Companion of the 13th international conference on {Software} engineering  - {ICSE} {Companion} '08},
	publisher = {ACM Press},
	author = {Cottrell, Rylan and Walker, Robert J. and Denzinger, Jörg},
	year = {2008},
	pages = {933},
}

@inproceedings{Gharehyazie2017,
	address = {Buenos Aires, Argentina},
	title = {Some from {Here}, {Some} from {There}: {Cross}-{Project} {Code} {Reuse} in {GitHub}},
	isbn = {978-1-5386-1544-7},
	shorttitle = {Some from {Here}, {Some} from {There}},
	url = {http://ieeexplore.ieee.org/document/7962379/},
	doi = {10.1109/MSR.2017.15},
	language = {en},
	urldate = {2022-05-03},
	booktitle = {2017 {IEEE}/{ACM} 14th {International} {Conference} on {Mining} {Software} {Repositories} ({MSR})},
	publisher = {IEEE},
	author = {Gharehyazie, Mohammad and Ray, Baishakhi and Filkov, Vladimir},
	month = may,
	year = {2017},
	pages = {291--301},
}

@misc{Lyu2024,
	title = {Automatic {Programming}: {Large} {Language} {Models} and {Beyond}},
	shorttitle = {Automatic {Programming}},
	url = {http://arxiv.org/abs/2405.02213},
	language = {en},
	urldate = {2024-08-21},
	publisher = {arXiv},
	author = {Lyu, Michael R. and Ray, Baishakhi and Roychoudhury, Abhik and Tan, Shin Hwei and Thongtanunam, Patanamon},
	month = may,
	year = {2024},
	note = {arXiv:2405.02213 [cs]},
}

@misc{Jiang2024,
	title = {A {Survey} on {Large} {Language} {Models} for {Code} {Generation}},
	url = {http://arxiv.org/abs/2406.00515},
	language = {en},
	urldate = {2024-06-17},
	publisher = {arXiv},
	author = {Jiang, Juyong and Wang, Fan and Shen, Jiasi and Kim, Sungju and Kim, Sunghun},
	month = jun,
	year = {2024},
	note = {arXiv:2406.00515 [cs]},
}

@article{Dong2024selfcollaboration,
	title = {Self-collaboration {Code} {Generation} via {ChatGPT}},
	issn = {1049-331X, 1557-7392},
	url = {https://dl.acm.org/doi/10.1145/3672459},
	doi = {10.1145/3672459},
	language = {en},
	urldate = {2024-07-04},
	journal = {ACM Transactions on Software Engineering and Methodology},
	author = {Dong, Yihong and Jiang, Xue and Jin, Zhi and Li, Ge},
	month = jun,
	year = {2024},
	pages = {3672459},
}

@misc{Xia2023,
	title = {Keep the {Conversation} {Going}: {Fixing} 162 out of 337 bugs for \$0.42 each using {ChatGPT}},
	shorttitle = {Keep the {Conversation} {Going}},
	url = {http://arxiv.org/abs/2304.00385},
	language = {en},
	urldate = {2023-08-20},
	publisher = {arXiv},
	author = {Xia, Chunqiu Steven and Zhang, Lingming},
	month = apr,
	year = {2023},
	note = {arXiv:2304.00385 [cs]},
}

@inproceedings{Wei2022,
author = {Wei, Jason and Wang, Xuezhi and Schuurmans, Dale and Bosma, Maarten and Ichter, Brian and Xia, Fei and Chi, Ed H. and Le, Quoc V. and Zhou, Denny},
title = {Chain-of-thought prompting elicits reasoning in large language models},
year = {2024},
isbn = {9781713871088},
publisher = {Curran Associates Inc.},
address = {Red Hook, NY, USA},
booktitle = {Proceedings of the 36th International Conference on Neural Information Processing Systems},
articleno = {1800},
numpages = {14},
location = {New Orleans, LA, USA},
series = {NIPS '22}
}

@article{Yang2024,
  title={Revisiting Unnaturalness for Automated Program Repair in the Era of Large Language Models},
  author={Yang, Aidan ZH and Kolak, Sophia and Hellendoorn, Vincent J and Martins, Ruben and Goues, Claire Le},
  journal={arXiv preprint arXiv:2404.15236},
  year={2024}
}

@inproceedings{Sajnani2016,
	address = {Austin Texas},
	title = {{SourcererCC}: scaling code clone detection to big-code},
	isbn = {978-1-4503-3900-1},
	shorttitle = {{SourcererCC}},
	url = {https://dl.acm.org/doi/10.1145/2884781.2884877},
	doi = {10.1145/2884781.2884877},
	language = {en},
	urldate = {2022-03-15},
	booktitle = {Proceedings of the 38th {International} {Conference} on {Software} {Engineering}},
	publisher = {ACM},
	author = {Sajnani, Hitesh and Saini, Vaibhav and Svajlenko, Jeffrey and Roy, Chanchal K. and Lopes, Cristina V.},
	month = may,
	year = {2016},
	pages = {1157--1168},
}

@misc{Yang2023,
	title = {Deep {Learning} {Based} {Code} {Generation} {Methods}: {A} {Literature} {Review}},
	shorttitle = {Deep {Learning} {Based} {Code} {Generation} {Methods}},
	url = {http://arxiv.org/abs/2303.01056},
	urldate = {2023-03-12},
	publisher = {arXiv},
	author = {Yang, Zezhou and Chen, Sirong and Gao, Cuiyun and Li, Zhenhao and Li, Ge and Lv, Rongcong},
	month = mar,
	year = {2023},
	note = {arXiv:2303.01056 [cs]},
}

@misc{Zhang2023Survey,
      title={A Survey of Learning-based Automated Program Repair}, 
      author={Quanjun Zhang and Chunrong Fang and Yuxiang Ma and Weisong Sun and Zhenyu Chen},
      year={2023},
      eprint={2301.03270},
      archivePrefix={arXiv},
      primaryClass={cs.SE},
      url={https://arxiv.org/abs/2301.03270}, 
}

@inproceedings{Eladawy2024,
author = {Eladawy, Hadeel and Le Goues, Claire and Brun, Yuriy},
title = {Automated Program Repair, What Is It Good For? Not Absolutely Nothing!},
year = {2024},
isbn = {9798400702174},
publisher = {Association for Computing Machinery},
address = {New York, NY, USA},
url = {https://doi.org/10.1145/3597503.3639095},
doi = {10.1145/3597503.3639095},
booktitle = {Proceedings of the IEEE/ACM 46th International Conference on Software Engineering},
articleno = {84},
numpages = {13},
location = {Lisbon, Portugal},
series = {ICSE '24}
}

@misc{Ouyang2023,
      title={LLM is Like a Box of Chocolates: the Non-determinism of ChatGPT in Code Generation}, 
      author={Shuyin Ouyang and Jie M. Zhang and Mark Harman and Meng Wang},
      year={2023},
      eprint={2308.02828},
      archivePrefix={arXiv},
      primaryClass={cs.SE},
      url={https://arxiv.org/abs/2308.02828}, 
}

@misc{Doderlein2023,
      title={Piloting Copilot and Codex: Hot Temperature, Cold Prompts, or Black Magic?}, 
      author={Jean-Baptiste Döderlein and Mathieu Acher and Djamel Eddine Khelladi and Benoit Combemale},
      year={2023},
      eprint={2210.14699},
      archivePrefix={arXiv},
      primaryClass={cs.SE},
      url={https://arxiv.org/abs/2210.14699}, 
}

@misc{Zhao2023,
	title = {A {Survey} of {Large} {Language} {Models}},
	url = {http://arxiv.org/abs/2303.18223},
	language = {en},
	urldate = {2023-08-20},
	publisher = {arXiv},
	author = {Zhao, Wayne Xin and Zhou, Kun and Li, Junyi and Tang, Tianyi and Wang, Xiaolei and Hou, Yupeng and Min, Yingqian and Zhang, Beichen and Zhang, Junjie and Dong, Zican and Du, Yifan and Yang, Chen and Chen, Yushuo and Chen, Zhipeng and Jiang, Jinhao and Ren, Ruiyang and Li, Yifan and Tang, Xinyu and Liu, Zikang and Liu, Peiyu and Nie, Jian-Yun and Wen, Ji-Rong},
	month = jun,
	year = {2023},
	note = {arXiv:2303.18223 [cs]},
}

@article{Brandt2009,
	title = {Writing {Code} to {Prototype}, {Ideate}, and {Discover}},
	volume = {26},
	doi = {10.1109/MS.2009.147},
	number = {5},
	journal = {IEEE Software},
	author = {Brandt, Joel and Guo, Philip J. and Lewenstein, Joel and Dontcheva, Mira and Klemmer, Scott R.},
	year = {2009},
	pages = {18--24},
}

@inproceedings{Srinath2017,
  title={Python – The Fastest Growing Programming Language},
  author={K. R. Srinath},
  year={2017},
  url={https://api.semanticscholar.org/CorpusID:218773526}
}

@online{GitHubAPI,
  author={GitHub},
  title={GitHub REST API},
  year=2022,
  note = {Accessed: 2024-04},
  url={https://docs.github.com/en/rest/search/search?apiVersion=2022-11-28},
}

@online{StackAPI,
  author={Stack},
  title={Exchange API v2.3},
  year=2021,
  note = {Accessed: 2024-04},
  url={https://api.stackexchange.com/docs},
}

@online{TreeSitter,
  author={tree-sitter},
  title={Tree-sitter},
  year=2024,
  note = {Accessed: 2024-05},
  url={https://tree-sitter.github.io/tree-sitter/},
}

@article{Braun2006,
    author = { Virginia   Braun  and  Victoria   Clarke },
    title = {Using thematic analysis in psychology},
    journal = {Qualitative Research in Psychology},
    volume = {3},
    number = {2},
    pages = {77-101},
    year  = {2006},
    publisher = {Routledge},
    doi = {10.1191/1478088706qp063oa},
    URL = {https://www.tandfonline.com/doi/abs/10.1191/1478088706qp063oa},
    eprint = {https://www.tandfonline.com/doi/pdf/10.1191/1478088706qp063oa}
}

@BOOK{Vladimir2020,
  author={Khorikov, Vladimir},
  booktitle={Unit Testing Principles, Practices, and Patterns},
  year={2020},
  volume={},
  number={},
  pages={},
  doi={}}

@inproceedings{Huang2018,
	address = {Montpellier France},
	title = {{ClDiff}: generating concise linked code differences},
	isbn = {978-1-4503-5937-5},
	shorttitle = {{ClDiff}},
	url = {https://dl.acm.org/doi/10.1145/3238147.3238219},
	doi = {10.1145/3238147.3238219},
	language = {en},
	urldate = {2024-01-20},
	booktitle = {Proceedings of the 33rd {ACM}/{IEEE} {International} {Conference} on {Automated} {Software} {Engineering}},
	publisher = {ACM},
	author = {Huang, Kaifeng and Chen, Bihuan and Peng, Xin and Zhou, Daihong and Wang, Ying and Liu, Yang and Zhao, Wenyun},
	month = sep,
	year = {2018},
	pages = {679--690},
}

@INPROCEEDINGS{Zhang2025IorI,
  author={Zhang, Tanghaoran and Yu, Yue and Mao, Xinjun and Wang, Shangwen and Yang, Kang and Lu, Yao and Zhang, Zhang and Zhao, Yuxin},
  booktitle={2025 IEEE/ACM 47th International Conference on Software Engineering (ICSE)}, 
  title={Instruct or Interact? Exploring and Eliciting LLMs' Capability in Code Snippet Adaptation Through Prompt Engineering}, 
  year={2025},
  volume={},
  number={},
  pages={566-577},
  doi={10.1109/ICSE55347.2025.00104}}

@misc{qwq32b,
    title = {QwQ-32B: Embracing the Power of Reinforcement Learning},
    url = {https://qwenlm.github.io/blog/qwq-32b/},
    author = {Qwen Team},
    month = {March},
    year = {2025}
}

@article{hui2024qwen2.5,
  title={Qwen2. 5-Coder Technical Report},
  author={Hui, Binyuan and Yang, Jian and Cui, Zeyu and Yang, Jiaxi and Liu, Dayiheng and Zhang, Lei and Liu, Tianyu and Zhang, Jiajun and Yu, Bowen and Dang, Kai and others},
  journal={arXiv preprint arXiv:2409.12186},
  year={2024}
}

@misc{deepseek-ai2024v3,
      title={DeepSeek-V3 Technical Report}, 
      author={DeepSeek-AI},
      year={2024},
      eprint={2412.19437},
      archivePrefix={arXiv},
      primaryClass={cs.CL},
      url={https://arxiv.org/abs/2412.19437}, 
}

@misc{deepseek-ai2025r1,
	title = {{DeepSeek}-{R1}: {Incentivizing} {Reasoning} {Capability} in {LLMs} via {Reinforcement} {Learning}},
	shorttitle = {{DeepSeek}-{R1}},
	url = {http://arxiv.org/abs/2501.12948},
	doi = {10.48550/arXiv.2501.12948},
	urldate = {2025-03-12},
	publisher = {arXiv},
	author = {DeepSeek-AI and Guo, Daya and Yang, Dejian and Zhang, Haowei and Song, Junxiao and Zhang, Ruoyu and Xu, Runxin and Zhu, Qihao and Ma, Shirong and Wang, Peiyi and others},
	month = jan,
	year = {2025},
	note = {arXiv:2501.12948 [cs]},
}

@misc{openai2024o1,
	title = {{OpenAI} o1 {System} {Card}},
	url = {http://arxiv.org/abs/2412.16720},
	doi = {10.48550/arXiv.2412.16720},
	urldate = {2025-03-12},
	publisher = {arXiv},
	author = {OpenAI and Jaech, Aaron and Kalai, Adam and Lerer, Adam and Richardson, Adam and El-Kishky, Ahmed and Low, Aiden and Helyar, Alec and Madry, Aleksander and Beutel, Alex and others},
	month = dec,
	year = {2024},
	note = {arXiv:2412.16720 [cs]},
}

@misc{jain2024LiveCodeBench,
      title={LiveCodeBench: Holistic and Contamination Free Evaluation of Large Language Models for Code}, 
      author={Naman Jain and King Han and Alex Gu and Wen-Ding Li and Fanjia Yan and Tianjun Zhang and Sida Wang and Armando Solar-Lezama and Koushik Sen and Ion Stoica},
      year={2024},
      eprint={2403.07974},
      archivePrefix={arXiv},
      primaryClass={cs.SE},
      url={https://arxiv.org/abs/2403.07974}, 
}

@inproceedings{Wang2024oop,
    title = "{OOP}: Object-Oriented Programming Evaluation Benchmark for Large Language Models",
    author = "Wang, Shuai  and
      Ding, Liang  and
      Shen, Li  and
      Luo, Yong  and
      Du, Bo  and
      Tao, Dacheng",
    editor = "Ku, Lun-Wei  and
      Martins, Andre  and
      Srikumar, Vivek",
    booktitle = "Findings of the Association for Computational Linguistics: ACL 2024",
    month = aug,
    year = "2024",
    address = "Bangkok, Thailand",
    publisher = "Association for Computational Linguistics",
    url = "https://aclanthology.org/2024.findings-acl.808/",
    doi = "10.18653/v1/2024.findings-acl.808",
    pages = "13619--13639"
}

@inproceedings{Cao2024JavaBench,
address = {Sacramento CA USA},
title = {{JavaBench}: {A} {Benchmark} of {Object}-{Oriented} {Code} {Generation} for {Evaluating} {Large} {Language} {Models}},
isbn = {979-8-4007-1248-7},
shorttitle = {{JavaBench}},
url = {https://dl.acm.org/doi/10.1145/3691620.3695470},
doi = {10.1145/3691620.3695470},
language = {en},
urldate = {2025-05-28},
booktitle = {Proceedings of the 39th {IEEE}/{ACM} {International} {Conference} on {Automated} {Software} {Engineering}},
publisher = {ACM},
author = {Cao, Jialun and Chen, Zhiyong and Wu, Jiarong and Cheung, Shing-Chi and Xu, Chang},
month = oct,
year = {2024},
pages = {870--882},
}

@misc{Li2024EvoCodeBench,
title = {{EvoCodeBench}: {An} {Evolving} {Code} {Generation} {Benchmark} {Aligned} with {Real}-{World} {Code} {Repositories}},
shorttitle = {{EvoCodeBench}},
url = {http://arxiv.org/abs/2404.00599},
language = {en},
urldate = {2024-04-18},
publisher = {arXiv},
author = {Li, Jia and Li, Ge and Zhang, Xuanming and Dong, Yihong and Jin, Zhi},
month = mar,
year = {2024},
note = {arXiv:2404.00599 [cs]},
}

@inproceedings{Feng2024Complex,
address = {Sacramento CA USA},
title = {{ComplexCodeEval}: {A} {Benchmark} for {Evaluating} {Large} {Code} {Models} on {More} {Complex} {Code}},
isbn = {979-8-4007-1248-7},
shorttitle = {{ComplexCodeEval}},
url = {https://dl.acm.org/doi/10.1145/3691620.3695552},
doi = {10.1145/3691620.3695552},
language = {en},
urldate = {2025-05-28},
booktitle = {Proceedings of the 39th {IEEE}/{ACM} {International} {Conference} on {Automated} {Software} {Engineering}},
publisher = {ACM},
author = {Feng, Jia and Liu, Jiachen and Gao, Cuiyun and Chong, Chun Yong and Wang, Chaozheng and Gao, Shan and Xia, Xin},
month = oct,
year = {2024},
pages = {1895--1906},
}

@inproceedings{Holtzman2019,
title = {The {Curious} {Case} of {Neural} {Text} {Degeneration}},
url = {https://openreview.net/forum?id=rygGQyrFvH},
language = {en},
urldate = {2025-04-09},
author = {Holtzman, Ari and Buys, Jan and Du, Li and Forbes, Maxwell and Choi, Yejin},
month = sep,
year = {2019}
}

@article{Cohen1960,
  title={A Coefficient of Agreement for Nominal Scales},
  author={Jacob Cohen},
  journal={Educational and Psychological Measurement},
  year={1960},
  volume={20},
  pages={37 - 46}
}

@misc{Li2025,
	title = {When {Thinking} {Fails}: {The} {Pitfalls} of {Reasoning} for {Instruction}-{Following} in {LLMs}},
	shorttitle = {When {Thinking} {Fails}},
	url = {http://arxiv.org/abs/2505.11423},
	doi = {10.48550/arXiv.2505.11423},
	urldate = {2025-07-25},
	publisher = {arXiv},
	author = {Li, Xiaomin and Yu, Zhou and Zhang, Zhiwei and Chen, Xupeng and Zhang, Ziji and Zhuang, Yingying and Sadagopan, Narayanan and Beniwal, Anurag},
	month = may,
	year = {2025},
	note = {arXiv:2505.11423 [cs]},
}

@misc{Fu2025,
	title = {Scaling {Reasoning}, {Losing} {Control}: {Evaluating} {Instruction} {Following} in {Large} {Reasoning} {Models}},
	shorttitle = {Scaling {Reasoning}, {Losing} {Control}},
	url = {http://arxiv.org/abs/2505.14810},
	doi = {10.48550/arXiv.2505.14810},
	urldate = {2025-07-25},
	publisher = {arXiv},
	author = {Fu, Tingchen and Gu, Jiawei and Li, Yafu and Qu, Xiaoye and Cheng, Yu},
	month = may,
	year = {2025},
	note = {arXiv:2505.14810 [cs]},
}

@inproceedings{Dong2024,
	address = {Bangkok, Thailand and virtual meeting},
	title = {Generalization or {Memorization}: {Data} {Contamination} and {Trustworthy} {Evaluation} for {Large} {Language} {Models}},
	shorttitle = {Generalization or {Memorization}},
	url = {https://aclanthology.org/2024.findings-acl.716},
	doi = {10.18653/v1/2024.findings-acl.716},
	language = {en},
	urldate = {2025-08-15},
	booktitle = {Findings of the {Association} for {Computational} {Linguistics} {ACL} 2024},
	publisher = {Association for Computational Linguistics},
	author = {Dong, Yihong and Jiang, Xue and Liu, Huanyu and Jin, Zhi and Gu, Bin and Yang, Mengfei and Li, Ge},
	year = {2024},
	pages = {12039--12050},
}

@INPROCEEDINGS{Zhang2025Unseen,
  author={Zhang, Yuanliang and Xie, Yifan and Lit, Shanshan and Liu, Ke and Wang, Chong and Jia, Zhouyang and Huang, Xiangbing and Song, Jie and Luo, Chaopeng and Zheng, Zhizheng and Xu, Rulin and Liu, Yitong and Zheng, Si and Liao, Xiangke},
  booktitle={2025 IEEE/ACM 47th International Conference on Software Engineering (ICSE)}, 
  title={Unseen Horizons: Unveiling the Real Capability of LLM Code Generation Beyond the Familiar}, 
  year={2025},
  volume={},
  number={},
  pages={604-615},
  doi={10.1109/ICSE55347.2025.00082}}

@article{Kong2025,
	title = {Demystifying {Memorization} in {LLM}-{Based} {Program} {Repair} via a {General} {Hypothesis} {Testing} {Framework}},
	volume = {2},
	issn = {2994-970X},
	url = {https://dl.acm.org/doi/10.1145/3729390},
	doi = {10.1145/3729390},
	language = {en},
	number = {FSE},
	urldate = {2025-07-10},
	journal = {Proceedings of the ACM on Software Engineering},
	author = {Kong, Jiaolong and Xie, Xiaofei and Liu, Shangqing},
	month = jun,
	year = {2025},
	note = {Publisher: Association for Computing Machinery (ACM)},
	pages = {2712--2734},
}

@misc{Reux2025,
      title={LLM Code Customization with Visual Results: A Benchmark on TikZ}, 
      author={Charly Reux and Mathieu Acher and Djamel Eddine Khelladi and Olivier Barais and Clément Quinton},
      year={2025},
      eprint={2505.04670},
      archivePrefix={arXiv},
      primaryClass={cs.SE},
      url={https://arxiv.org/abs/2505.04670}, 
}

@misc{Liu2021,
	title = {Pre-train, {Prompt}, and {Predict}: {A} {Systematic} {Survey} of {Prompting} {Methods} in {Natural} {Language} {Processing}},
	shorttitle = {Pre-train, {Prompt}, and {Predict}},
	url = {http://arxiv.org/abs/2107.13586},
	language = {en},
	urldate = {2023-08-21},
	publisher = {arXiv},
	author = {Liu, Pengfei and Yuan, Weizhe and Fu, Jinlan and Jiang, Zhengbao and Hayashi, Hiroaki and Neubig, Graham},
	month = jul,
	year = {2021},
	note = {arXiv:2107.13586 [cs]},
}

\end{document}